\documentclass{JHEP3} 

\usepackage{epsfig}
\usepackage{amsmath, amssymb}
\usepackage{concmath, palatino}
\usepackage{mathrsfs}
\usepackage{color}

\newcommand\fverb{\setbox\pippobox=\hbox\bgroup\verb}
\newcommand\fverbdo{\egroup\medskip\noindent%
			\fbox{\unhbox\pippobox}\ }
\newcommand\fverbit{\egroup\item[\fbox{\unhbox\pippobox}]}
\newbox\pippobox

\newcommand{\be}{\begin{equation}}
\newcommand{\ee}{\end{equation}}
\newcommand{\ba}{\begin{eqnarray}}
\newcommand{\ea}{\end{eqnarray}}

\newcommand{\refeq}[1]{Eq.~(\ref{eq:#1})}

\newcommand{\ads}{AdS_5\times S^5}

\newcommand{\ddb}{{\overline{\mathscr D}}}
\newcommand{\sym}{$\mathcal{N}=4$ SYM }

\allowdisplaybreaks

    \renewcommand{\(}{\left(}
    \renewcommand{\)}{\right)}
    \newcommand{\eq}[1]{(\ref{#1})}

    \newcommand{\beq}{\begin{equation}}
    \newcommand{\eeq}{\end{equation}}
    \newcommand\beqa{\begin{eqnarray}}
    \newcommand\eeqa{\end{eqnarray}}
        \renewcommand{\d}{\partial}

\title{On wrapping corrections to GKP-like operators}

\author{Matteo Beccaria\\
  Dipartimento di Fisica, Universita' del Salento, 
  Via Arnesano, 73100 Lecce \&\\
  INFN, Sezione di Lecce\\
  E-mail: \email{matteo.beccaria$\bullet$le.infn.it}
}

\author{Fedor~Levkovich-Maslyuk\\ Physics Department, Moscow State University, 119991, Moscow, Russia\\
    E-mail: \email{fedor.levkovich$\bullet$gmail.com}}

\author{Guido Macorini\\
  Dipartimento di Fisica, Universita' del Salento, 
  Via Arnesano, 73100 Lecce \&\\
  INFN, Sezione di Lecce\\
  E-mail: \email{guido.macorini$\bullet$le.infn.it}
}

\abstract{
In the recent paper {\tt arXiv:1010.5009}, Maldacena {\em et al.} derive the two loop expressions for polygonal 
Wilson loops expectation values, or MHV amplitudes, by writing them as sums
over exchanges of  intermediate free particles. The spectrum of excitations of the flux tube between two null Wilson lines  can be viewed as the spectrum of excitations around the infinite spin limit of finite twist 
operators in the $\mathfrak{sl}(2)$ sector of \sym, or the Gubser-Klebanov-Polyakov (GKP) string. This regime 
can be captured exploiting integrability and assuming that wrapping corrections are negligible compared to 
asymptotic Bethe Ansatz contributions. This assumption holds true for the \sym background GKP string, but deserves
further analysis for excited states. Here, we investigate GKP cousins by considering various classes of (generalized) twist operators in $\beta$-deformed \sym and ABJM theory. Assuming
that the Y-system of Gromov-Kazakov-Vieira correctly reproduces the wrapping corrections, we show that it
easily leads to  accurate large spin expansions at lowest order in weak-coupling perturbation theory. As a byproduct, 
we confirm these corrections are subleading
in all the considered cases.
}

\keywords{}
\preprint{}

\begin{document} 

\section{Introduction and motivations}
\label{sec:intro}

In the recent paper \cite{Gaiotto:2010fk}, Maldacena {\em et al.} derive the two loop expressions for polygonal 
Wilson loops expectation values, or MHV amplitudes, by starting from the one loop result and 
applying an operator product expansion. The various terms in the OPE are associated with the 
exchange of free particles. 
The inclusion of the one loop energy/anomalous dimension of each intermediate particle breaks the
cyclic symmetry of the amplitude. In \cite{Gaiotto:2010fk}, it is shown that the simplest cyclic completion agrees
 with explicit results computed by more direct methods \cite{DelDuca:2010zp}.
The spectrum of excitations of the flux tube stretching between two null Wilson lines
 can also be viewed as the spectrum of excitations around the infinite spin limit of finite twist 
 operators in the $\mathfrak{sl}(2)$ sector of \sym, 
 or the GKP  \cite{Gubser:2002tv} string.

Integrability and AdS/CFT correspondence effectively help in computing such spectrum. To this aim, the 
flux tube is mapped to the GKP state which is the $\mathfrak{sl}(2)$ \sym operator ~\footnote{
As usual, dots in 
Eq.~(\ref{eq:GKP}) stand for a suitable combination of fields and derivatives inside the trace building an exact eigenstate of the dilatation operator.
}
\be
\label{eq:GKP}
\mathcal{O}_{\rm GKP} = \mbox{Tr}(Z\,\, D^{N}_{+}\,Z)+\dots\,.
\ee
Here, $D_{+}$ is a light-cone direction and $Z$ is one of the complex scalars of \sym.  According to the approach
of \cite{Gaiotto:2010fk}, the two scalars can be regarded as two fast particles sourcing the flux tube 
represented by the light-cone derivatives. Finally, the spin $N$ is to be taken to infinity. 
Excitations over the GKP string are more involved and are associated with operators of the form 
\be
\label{eq:excited-GKP}
\mathcal{O}_{\rm GKP}^{*} = \mbox{Tr}(Z\,\, D^{N_{1}}_{+}\,\chi\,D_{+}^{N_{2}}\,Z)+\dots\,.
\ee
where the field $\chi$ moves in the background of derivatives. Again, the computation of the anomalous dimensions
of the excited states $\mathcal{O}_{\rm GKP}^{*}$ is viable thanks to integrability as discussed in 
details in \cite{Basso:2010in}. The treatment is based on the continuum, large spin limit
of the asymptotic Bethe Ansatz equations \cite{ABA}
of \sym.
%

However,  it is well known that these equations do not capture correctly the interactions that wrap 
around the spin chain \cite{Wrapping}. 
In general, the size of these corrections is correlated to the spin-chain length and becomes more important 
for short length. It remains to be seen whether and to what extent wrapping corrections are subleading in the large
spin limit. The common lore is that wrapping issues are circumvented in the large spin limit. For instance, 
this claim is well supported in the case of the GKP background from explicit 
computations of wrapping corrections of twist-2~\footnote{With similar results
holding for twist-3 operators.} anomalous dimensions in the $\mathfrak{sl}(2)$ 
sector \cite{Wrapping-subleading,Bajnok:2008qj}. 
These results indicate that wrapping corrections are of order $\mathcal{O}(\log^{2} N/N^{2})$. 
A more careful analysis as well as extensions would entail the use of the Y-system of \cite{GKV}, under the 
assumption that it correctly computes wrapping effects~\footnote{This is a reasonable assumption, but it must be stressed that comparisons with explicit results in $\beta$-deformed theories and ABJM should be welcome.}.

Actually, extensions can go in two directions. Of course, 
the most ambitious goal is precisely that of working out the large
spin wrapping for operators like $\mathcal{O}_{\rm GKP}^{*}$. In this paper, we start from a simpler problem
in order to develop useful methods and tools. We consider two \sym cousins, {\em i.e.}
integrable $\beta$ deformations  \cite{LeiStr,MPSZ} and ABJM theory  \cite{Aharony:2008ug}, and 
analyze wrapping for GKP-like states. These are associated by duality to generalized GKP operators or, more
briefly, {\em twist operators}. They have the characteristic feature of being built by adding to a certain number of elementary fields a {\em sea}
of $N$ covariant derivatives. For such twist operators, we shall show how to extract from the $Y$-system, the large $N$
expansion of the leading weak-coupling wrapping correction (see footnote 3). We shall focus on 
operators built in $\mathfrak{sl}(2)$-like sectors. In all cases, 
we shall be able to provide accurate asymptotic expansions of the wrapping correction proving that it is indeed subleading
compared to the leading logarithmic scaling of anomalous dimensions \cite{logarithmic-scaling}. 
The obtained expansions 
nicely complement analogous large spin results derived for the asymptotic Bethe Ansatz contributions
in \cite{Catino:2009zz}~\footnote{
Notice that the large spin limit can be taken at the level of the ABA equations. In that context, the density of Bethe roots
is obtained by a continuum limit turning the Bethe equations into integral equations~\cite{IntegralEquations}. This method is more general 
than the approach of \cite{Catino:2009zz}, but is currently unable to provide long expansions at large spin and is 
limited to the leading and subleading terms.}. 
One immediate application of these results is the analysis of generalized
Gribov-Lipatov reciprocity (see \cite{reciprocity-review} for a review). For instance, we shall provide 
support for reciprocity to hold in the case of $\beta$ deformed \sym.

\section{A brief review of Y-system for undeformed $\mathcal{N}=4$ SYM}
\label{sec:Y-system-undeformed}

Superstring theory of type IIB on $\ads$, described by the Metsaev-Tseytlin action in light-cone gauge, is a two dimensional classically integrable field theory. According to the AdS/CFT correspondence, it
is dual to four dimensional planar $\mathcal{N}=4$ SYM. The spectrum of anomalous dimensions of the gauge theory is believed 
to match the superstring energy spectrum.

\medskip
The exact spectrum of relativistic 2D integrable theories has been suggested~\cite{Zamolodchikov:1991et} 
to be captured by the universal set of functional quadratic
Hirota equations. They have the form
\be
\label{Tsystem}
  T_{a,s}(u+i/2)\,T_{a,s}(u-i/2) =T_{a+1,s}(u)\,T_{a-1,s}(u)+T_{a,s+1}(u)\,T_{a,s-1}(u)\;.
\ee
It was conjectured in \cite{GKV} that for the superstring theory on $\ads$
 the system of Hirota equations
should be the same, with the functions $T_{a,s}(u)$ being non-zero
only inside the infinite T-shaped domain of the ${a,s}$ integer lattice, shown in Fig.\ref{Fig:figysys}.

\FIGURE[ht]
{\label{Fig:figysys}

    \begin{tabular}{cc}
    \includegraphics[scale=0.6]{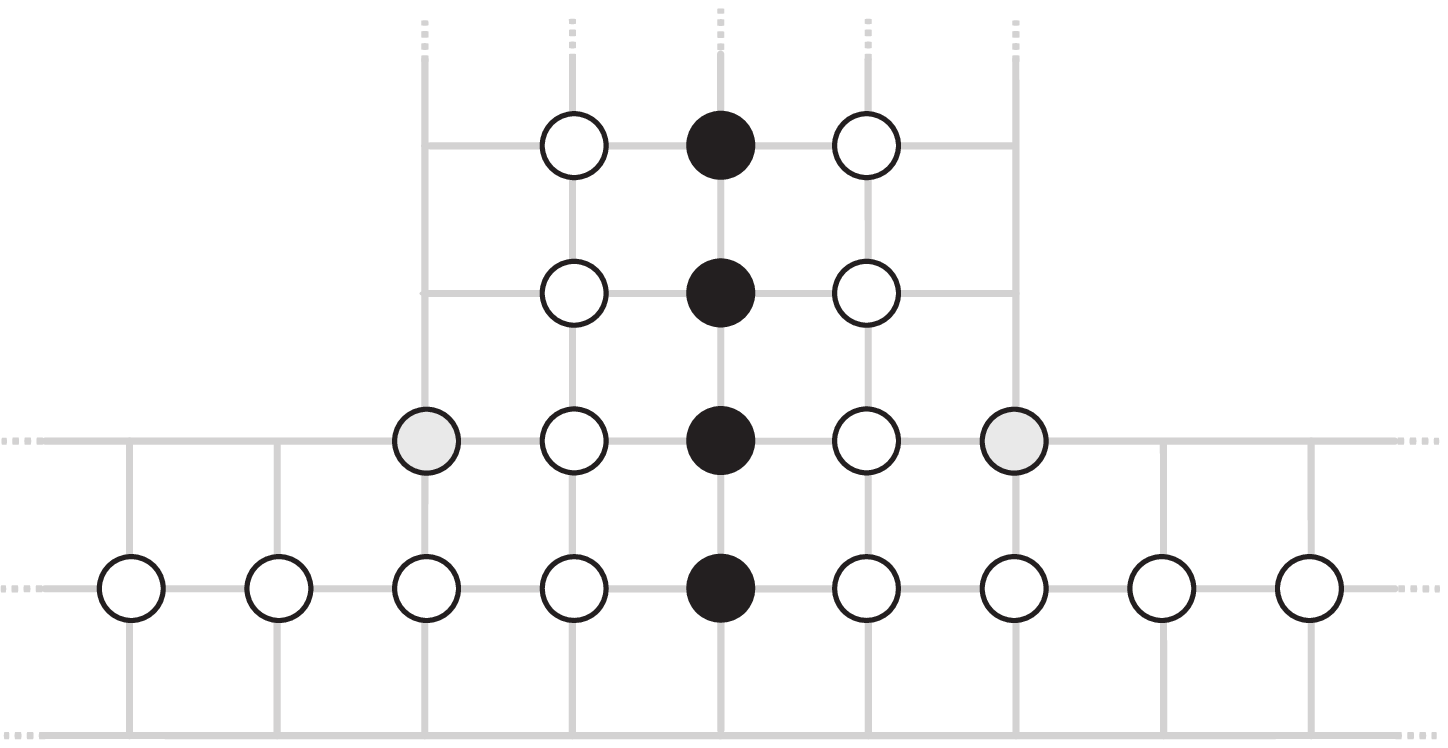}
    \end{tabular}
    \caption{Graphical representation of the Y-system and T-system \cite{GKV}.
    Circles
    correspond to Y-functions. Intersections of grid lines in the T-hook correspond to T-functions.}
}

Physical quantities can be computed by introducing the gauge invariant combinations
\begin{equation}
\label{YT}
Y_{a,s}=\frac{T_{a,s+1}T_{a,s-1}}{T_{a+1,s}T_{a-1,s}}\;\; ,
\end{equation}
which obey another set of functional equations called the Y-system:\footnote{The equations for $\{ a,s\} =\{2, 2\}$ and $\{a,s\} =\{-2, 2\}$ cannot be written in such local form.}
\begin{equation}
\label{eq:Ysystem}
\frac{Y_{a,s}^+ Y_{a,s}^-}{Y_{a+1,s}Y_{a-1,s}}
  =\frac{(1+Y_{a,s+1})(1+Y_{a,s-1})}{(1+Y_{a+1,s})(1+Y_{a-1,s})} \, ,
\end{equation}
Here and throughout this paper we use the notation
\be
	f^\pm \equiv f(u\pm i/2),\  f^{[+a]} \equiv f(u + ia/2).
\ee
The T- and Y-systems should additionally be supplemented with a particular set of analytical properties imposed on T- and Y-functions. What these properties are in the AdS/CFT case is still not completely clear  (for some recent progress see for instance \cite{Cavaglia:2010nm}). Also, recently the general solution of this Y-system was obtained \cite{Gromov:2010km}, though it is not clear at the moment what form it must take for any partcular operator/string state.

\medskip
The energy and momentum of magnon excitations in the theory are described
in terms of the
Zhukowski variable $x(u)$, defined by
    \be
    \label{xdef}
        x + \frac{1}{x} = \frac{u}{g},
    \ee
where the relation between the coupling $g$ and the 't Hooft coupling $\lambda$ is $\lambda = 16\pi^{2}g^{2}$.
The {\em mirror} and {\em physical} branches of this function are defined as
   \be
    \label{xBranches}
        x^{\rm ph}(u)=\frac{1}{2}\,\left(
        \frac{u}{g}+\sqrt{\frac{u}{g}-2}\;\sqrt{\frac{u}{g}+2}\right)\;\;,\;\;
        x^{\rm mir}(u)=\frac{1}{2}\left( \frac{u}{g}+i\sqrt{4-\frac{u^2}{g^2}}\right) \,,
    \ee
    where $\sqrt{u}$ stands for the principal branch of the square root. The energy and momentum of a bound state with $n$
    magnons are
 \begin{equation}
\label{epsandp}
\epsilon_n(u)= n+\frac{2ig}{x^{[+n]}}-\frac{2ig}{x^{[-n]}}\;\;,\;\;
p_n(u)=\frac{1}{i}\log\frac{x^{[+n]}}{x^{[-n]}}\;.
\end{equation}
Finally, the exact energy of a state was conjectured in \cite{GKV} to be given by
\be
\label{Efull}
E=\sum_{j}\epsilon_1^{\rm ph} (u_{4,j})+\delta E\;\;,\;\;
\delta E = \sum_{a=1}^\infty \int\frac{du}{2\pi i}
\,\,\frac{\partial\epsilon_a^{*} (u)}{\partial u} \log\left(1+Y_{a,0}^* (u)\right)\;,
\ee
where the rapidities $u_{4,j}$ are fixed by the exact Bethe ansatz equations
\be
Y_{1,0}^{\rm ph} (u_{4,j})=-1\,.
\ee
Here, $Y_{a,0}^*$ denotes the function $Y_{a,0}$ evaluated at mirror kinematics (and this applies for any function marked by asterix). On the other hand, $Y_{1,0}^{\rm ph}(u)$, similarly to $x^{\rm ph}$, is the result of analytical
continuation of $Y_{1,0}^{*}(u)$ through the cut $(i/2+2g, i/2+\infty)$.

\medskip
For asymptotically large size $L$ it is possible to solve the $Y$-system explicitly since the massive nodes 
$Y_{a,0}$ decouple and the Y-system splits into two wings $su_{\rm L}(2|2)\oplus su_{\rm R}(2|2)$. The solution found in this way can also be used to compute leading wrapping corrections to the anomalous dimensions at fixed finite $L$ \cite{GKV}.
Let us define for node $i$
\be
Q_{i}(u) = \prod_{\ell=1}^{K_{i}}(u-u_{i,\ell}),
\ee
and, for the momentum carrying node 4, let us introduce the quantities
\be
R^{(\pm)} =  \prod_{\ell=1}^{K_{4}}\left[x(u)-x^{\mp}_{4,\ell}\right], \qquad
B^{(\pm)} =  \prod_{\ell=1}^{K_{4}}\left[\frac{1}{x(u)}-x^{\mp}_{4,\ell}\right].
\ee
In the large size limit the $T_{a,1}^{\rm L}$ functions, which correspond to the left wing, can be found  from the following generating functional
\ba
\lefteqn{\sum_{a=0}^{\infty} (-1)^{a}\,T_{a,1}^{\rm L}\(u+i\tfrac{1-a}{2}\)\,\ddb^{a} = } && \\
&& 
\left(1-\frac{Q_{3}^{+}}{Q_{3}^{-}}\,\ddb\right)^{-1}\,
\left(1-\frac{Q_{3}^{+}}{Q_{3}^{-}}\frac{Q_{2}^{--}}{Q_{2}}\frac{R^{(+)-}}{R^{(-)-}}\,\ddb\right)
\left(1-\frac{Q_{2}^{++}}{Q_{2}}\frac{Q_{1}^{-}}{Q_{1}^{+}}\frac{R^{(+)-}}{R^{(-)-}}
\,\ddb\right)\,
\left(1-\frac{Q^{-}_{1}}{Q_{1}^{+}}\frac{B^{(+)+}}{B^{(-)+}}\frac{R^{(+)-}}{R^{(-)-}}\ddb\right)^{-1},
\nonumber
\ea
where $\ddb = e^{-i\partial_{u}}$ is the shift operator. For the right wing, we have a similar expression for the $T_{a,1}^{\rm R}$ functions with subscripts of functions $Q_i$ changed according to $1,2,3\to 7,6,5$. 

In this paper,
we shall consider states from the $sl(2)$ sector. In this case, there are excitations on the middle node only, which means that $K_i = 0$ for $i\neq 4$. Hence $Q_{1,2,3}\equiv 1$ and the T-functions for the left and right wing are equal: $T_{a,1}^{\rm L}=T_{a,1}^{\rm R}\equiv T_{a,1}$. They are given by
\ba
\lefteqn{\sum_{a=0}^{\infty} (-1)^{a}\,T_{a,1}\(u+i\tfrac{1-a}{2}\)\,\ddb^{a} = } && \\
&& 
\left(1-\ddb\right)^{-1}\,
\left(1-\frac{R^{(+)-}}{R^{(-)-}}\,\ddb\right)
\left(1-\frac{R^{(+)-}}{R^{(-)-}}
\,\ddb\right)\,
\left(1-\frac{B^{(+)+}}{B^{(-)+}}\frac{R^{(+)-}}{R^{(-)-}}\ddb\right)^{-1}.
\nonumber
\ea
The middle node Y-functions for large $L$ can then be written as
\be
\label{Yaasymp}
Y_{a,0}\simeq \(\frac{x^{[-a]}}{x^{[+a]}}\)^L \(T_{a,1}\)^2
\Phi_a\;,
\ee
where the fused scalar factor is 
\be
\label{phiin}
\Phi_a(u) = \prod_{n=-\frac{a-1}{2}}^{\frac{a-1}{2}}\Phi(u+i\,n), \qquad \Phi(u) = 
\frac{B^{(+)+}}{B^{(-)-}}\frac{R^{(-)-}}{R^{(+)+}}\prod_{j=1}^{K_4}\sigma^2(u,u_{4,j}).
\ee
and $\sigma (u,v)$ is the dressing factor.

In mirror dynamics $Y_{a,0}$ defined by \eq{Yaasymp} is suppressed for finite $L$ at weak coupling. The  wrapping correction at leading order in the coupling can be found from \eq{Efull}:
\be
\delta E^{\rm LO}\simeq -\sum_{a=1}^{\infty}\int\frac{du}{\pi} Y_{a,0}^*(u),
\ee
where values of the roots $u_{4,j}$ which enter the expression for $Y_{a,0}$ should be obtained from the standard ABA equations \cite{ABA} for the asymptotic spectrum of ${\mathcal N}=4$ SYM.

\subsection{$\mathfrak{sl}(2)$ twist operators}

An interesting class of composite operators in $\mathcal N=4$ SYM is that of twist $L$ operators with 
spin $N$. These are schematically of the form 
\be
\mbox{Tr}(D^{n_{1}}Z\,\cdots\, D^{n_{L}}Z), \qquad n_{1}+\cdots n_{L} = N,
\ee
where $D$ is a light-cone projected covariant derivative, and $Z$ a complex scalar field. The above set of operators
mixes under renormalization, but is closed at all orders in perturbation theory. These operators transforms according to the (reducible) infinite dimensional representation $[-\frac{1}{2}]^{\otimes L}$ of $\mathfrak{sl}(2)$.
We shall be interested in highest weight state with minimal anomalous dimension $\gamma_{L,N}(g)$. 
At weak-coupling, we can expand
\be
\gamma_{L,N}(g) = L+N+\sum_{\ell\ge 1} \gamma_{L, N}^{(\ell)}\,g^{2\,\ell}.
\ee
Wrapping corrections starts at order $g^{2\,L+2}$. The most advanced explicit results for the full (asymptotic plus wrapping) anomalous dimensions are at five loops for $L=2$ ~\cite{Lukowski:2009ce} and at six loops for $L=3$~\cite{Velizhanin:2010cm}. These results are in full agreement with BFKL as well as reciprocity~\cite{reciprocity-review} predictions.

\subsection{Efficient generation of $Y_{a,0}^{*}$ }

The quantity $Y_{a,0}^{*}$ is the product of the dispersion factor, the fused scalar factor, and the $T^{*}$-matrix factor, all evaluated in mirror dynamics.
We report here efficient formulas to compute these quantities for $\mathfrak{sl}(2)$ twist-operators. 

\medskip\noindent
\underline{\em Dispersion}

\medskip\noindent
This is the universal factor
\be
\left[\frac{4g^{2}}{(a^{2}+4u^{2})^{}}\right]^{L}.
\ee

\medskip\noindent
\underline{$su(2|2)$ \em Wing}

\medskip\noindent
We can use the relations
\ba
\frac{B^{(+)}}{B^{(-)}}\frac{R^{(+)}}{R^{(-)}} &=& \frac{Q_{4}^{+}}{Q_{4}^{-}}+{\cal O}(g^{4}), \\
\frac{R^{(+)}}{R^{(-)}} &=& \frac{Q_{4}^{+}}{Q_{4}^{-}}\,\left(1+i\,c\,g^{2}\,\frac{1}{u}
+{\cal O}(g^{4})\right),
\ea
where
\be
c = \sum_{j}\frac{1}{u_{4,j}+\frac{i}{2}}\frac{1}{u_{4,j}-\frac{i}{2}} = \left.
i\,(\log Q_{4})'\right|_{u=-i/2}^{u=+i/2}.
\ee
Then, one finds
\be
T_{a,0}^{*} = \left. i\,c\,g^{2}\,\frac{(-1)^{a+1}}{Q_{4}^{[1-a]}}
\mathop{\sum_{p=-a}^{a}}_{\Delta p = 2}\frac{Q_{4}^{[-1-p]}-Q_{4}^{[1-p]}}{u-p\frac{i}{2}} 
\right|_{Q_{4}^{[-a-1]} , Q_{4}^{[a+1]}\to 0}
\ee

\medskip\noindent
\underline{\em Fusion of scalar factors}

\medskip\noindent
After some manipulations, one finds (for an even $Q_{4}$) the formula
\be
\Phi^{*}_{a} = [(Q^{+}_{4}(0)]^{2} \,\frac{Q_{4}^{[1-a]}}{Q_{4}^{[-1-a]}Q_{4}^{[a-1]}Q_{4}^{[a+1]}}.
\ee

\subsection{Leading wrapping correction for twist-2 operators}

The leading wrapping correction for states with $L=2$ and spin $N$ has been derived in \cite{Bajnok:2008qj}
and reads
\ba
\label{eq:janik}
\gamma^{(4), \rm wrapping}_{2, N} &=& C_7(N) +C_4(N)\, \zeta_{3}+ C_2(N)\, \zeta_{5}, \\
C_2(N) &=& -640 S_1^2, \nonumber \\
C_4(N) &=& -512 S_1^2 S_{-2}, \nonumber \\
C_7(N) &=& 256 S_1^2 \left( -S_5+S_{-5}+2S_{4,1}-2S_{3,-2}+2S_{-2,-3}
-4S_{-2,-2,1} \right),\nonumber
\ea
where, as usual, harmonic sums are defined by 
\be
S_{a}(N) = \sum_{n=1}^{N}\frac{(-1)^{{\rm sign}\ a}}{n^{|a|}},\qquad
S_{a, \mathbf{b}}(N) = \sum_{n=1}^{N}\frac{(-1)^{{\rm sign}\ a}}{n^{|a|}}\,S_{\mathbf{b}}(n).
\ee
The Y-system calculation required the known one-loop Baxter polynomial for central roots which is 
\be
Q_{4}(u) = {}_{3}F_{2}\left(\left.\begin{array}{c}
-N\quad N+1\quad \frac{1}{2}+i\,u \\
1\quad 1
\end{array}\right| 1\right).
\ee

\subsubsection{A sample calculation, $N=2$ the Konishi operator}

As a warm-up, let us review the case of the Konishi operator which has $L=2$, $N=2$
and has already been discussed in \cite{GKV}. One finds
\ba
Y_{a,0}^{*} &=& \frac{147456 a^2 g^8 \left(3 a^2+12 u^2-4\right)^2}{\left(a^2+4 u^2\right)^4 
y_{a}y_{-a}}, \\
y_{a} &=& 9 a^4-36 a^3+72 a^2 u^2+60 a^2-144 a u^2-48 a+144 u^4+48 u^2+16. \nonumber
\ea
The residue in $u=i\frac{a}{2}$ is 
\ba
{\rm Res}_{u=i\frac{a}{2}} Y_{a,0}^{*} &=& -\frac{72 i P(a) g^8}{a^5 \left(9 a^4-3 a^2+1\right)^4}, \\
P(a) &=& 13122 a^{16}-63423 a^{14}+90396 a^{12}-52731 a^{10}+18792 a^8-4887 a^6+972 a^4-126 a^2+10.
\nonumber
\ea
Summing over $a$, we find 
\be
\gamma_{2, 2}^{(4), \rm wrapping}\,g^{8} = \sum_{a=1}^{\infty}(-2\,i){\rm Res}_{u=i\frac{a}{2}} Y_{a,0}^{*} =(324 +864\zeta_{3}-1440 \zeta_{5})\,g^{8},
\ee
in agreement with \refeq{janik}.

\subsubsection{Another example, $N=4$}

The same calculation can be repeated for the state with $N=4$. We provide some unpublished results
for the reader's advantage.
\ba
Y_{a,0}^{*} &=& \frac{655360000 a^2 g^8 \left(35 a^4-280 a^2 u^2+460 a^2-1680 u^4+1840 u^2-576\right)^2}{
9 \left(a^2+4 u^2\right)^4\,y_{a}y_{-a}}, \\
y_{a} &=& 1225 a^8+9800 a^7+19600 a^6 u^2+43400 a^6+117600 a^5 u^2+ \nonumber \\
&& 123200 a^5+117600 a^4 u^4+330400 a^4 u^2+241040 a^4+470400 a^3 u^4+537600 a^3 u^2+\nonumber \\
&& 325760 a^3+313600 a^2 u^6+560000 a^2 u^4+602240 a^2 u^2+290560 a^2+627200 a
   u^6+179200 a u^4+\nonumber \\
   &&442880 a u^2+153600 a+313600 u^8-268800 u^6+272640 u^4+117760 u^2+36864. \nonumber
\ea
The residue in $u=i\frac{a}{2}$ is 
\ba
{\rm Res}_{u=i\frac{a}{2}} Y_{a,0}^{*} &=& 
-\frac{12500\,i}{81 a^5 \left(35 a^4-70 a^3+85 a^2-50 a+12\right)^4 \left(35 a^4+70 a^3+85 a^2+50 a+12\right)^4}P(a), \nonumber\\
P(a) &=& 450375078125 a^{32}+5404500937500 a^{30}+12978155312500 a^{28}-22899012281250 a^{26}+\nonumber \\
&& -33135105543750 a^{24}-52511921568750 a^{22}+20889501517500 a^{20}+36484448411250 a^{18}+\nonumber \\
&& 57156710831625 a^{16}+10026729250
   a^{14}+1352770312800 a^{12}-10099371503200 a^{10}+\nonumber \\
   && 5545042560000 a^8-1729583746560 a^6+383012167680 a^4-51920289792 a^2+ \\
   && 3869835264.
\nonumber
\ea
Summing over $a$, we find 
\be
\gamma_{2, 4}^{(4), \rm wrapping}\,g^{8} = \sum_{a=1}^{\infty}(-2\,i){\rm Res}_{u=i\frac{a}{2}} Y_{a,0}^{*} =
\left(\frac{5196875}{7776}+\frac{143750 }{81}\,\zeta_{3}-\frac{25000}{9}\,\zeta_{5}\right)\,g^{8},
\ee
again in agreement with \refeq{janik}.

\subsubsection{General $N$}

In a similar way, one can easily to compute $\gamma^{(4), \rm wrapping}_{2, 4}$ for any
fixed $N$. Higher twist states are also similar and in all cases there is agreement with the known 
explicit results. Notice however that the proof of the exact closed-form dependence on $N$, as in \refeq{janik}, requires an educated guess.
This is available in the $\mathfrak{sl}(2)$ sector $\mathcal{N}=4$ SYM, but does not appear to be generalizable to larger sectors or models, at least in any trivial way.

\section{Y-system and $\beta$ deformation of $\mathcal{N}=4$ SYM}

A deformation of \sym which apparently does not spoil integrability in the planar limit is the so-called $\beta$-deformed SYM theory, which has ${\cal N}=1$ instead of ${\cal N}=4$ supersymmetry. In fact, this theory is a special case of a more general three-parameter deformation. The $\beta$-deformation consists in replacing the original superpotential for the chiral superfields by
\be
W= ih\ {\rm tr}
(e^{i\pi\beta}\phi\psi Z-e^{-i\pi\beta}\phi Z\psi).
\ee
The deformed theory remains superconformal in the planar limit to all orders of perturbation theory
\cite{LeiStr,MPSZ} if $\beta$
is real and $h{\overline h}=g_{\rm YM}^2$,
where $\lambda = g_{\rm YM}^{2}\,N$. 
The $\beta$-deformed theory is believed to have a string dual, and integrability properties have been found on both sides of the duality \cite{betarefs,ABBNs} (some important results in this field have been obtained quite recently). In particular, computations of wrapping corrections have been done with the use of integrability \cite{Gun,BecAng,ABBNw,Gromov:2010dy,Arbeta,Bajnok:2010ud}, reproducing the direct perturbative calculations obtained earlier in \cite{defdiagr} (see also the review \cite{defdiagrrev}).

The first application of the $Y$-system approach to $\beta$-deformed \sym has been presented in \cite{Gromov:2010dy}. In that paper it was argued that the deformed theory is described by the same Y-system as the undeformed one, though the asymptotic large $L$ solution should be modified and acquires dependence on $\beta$. Here we will use the expressions for this asymptotic solution in $sl(2)$ grading, which are given in the Appendix of \cite{Gromov:2010dy} and can be used for an arbitrary state. We will consider states from the $sl(2)$ sector, and in this case the $T_{a,s}^{\rm R,L}$ functions can be found from the following deformed generating functional:
\ba
\label{wbeta}
\lefteqn{\sum_{a=0}^{\infty} (-1)^{a}\,T_{a,1}^{\rm R}\(u+i\tfrac{1-a}{2}\)\,\ddb^{a} = } && \\
&& 
\left(1-\ddb\right)^{-1}\,
\left(1-\frac{1}{\lambda_{\rm R}}\frac{R^{(+)-}}{R^{(-)-}}\,\ddb\right)
\left(1-{\lambda_{\rm R}}\frac{R^{(+)-}}{R^{(-)-}}
\,\ddb\right)\,
\left(1-\frac{B^{(+)+}}{B^{(-)+}}\frac{R^{(+)-}}{R^{(-)-}}\ddb\right)^{-1}.
\nonumber
\ea
The expression for the $T_{a,1}^{\rm L}$ functions is the same with ${\lambda_{\rm R}}$ replaced by ${\lambda_{\rm L}}$. The left wing is undeformed, while deformation in the right wing is dependent on the length $L$:
\beq
	\ \lambda_{\rm L}=1, \ \lambda_{\rm R} = e^{2\pi i\beta L}.
\eeq 
The functional \eq{wbeta} leads to $T_{a,1}$ functions whose dependence on $\beta$ does not increase in complexity with $a$, which simplifies calculations considerably. Finally, the $Y_{a,0}$ functions are given by
\beq
	Y_{a,0}(u)\simeq \(\frac{x^{[-a]}}{x^{[+a]}}\)^L \Phi_a(u) T_{a,1}^{\rm L}(u) T_{a,1}^{\rm R}(u)
	\;,
\eeq
where the scalar factor $\Phi$ is the same as in the undeformed theory (see \eq{phiin}). The energy of a state is also given by the same formula \eq{Efull}.

\subsection{Deformed Konishi operator}
As we saw above, to compute $Y_{a,0}^*$ for deformed twist operators~\footnote{Let us remark that in the $\mathfrak{sl}(2)$ sector, the asymptotic Bethe Ansatz anomalous dimensions are independent on the twist which enters at the level of the wrapping corrections, see
for instance \cite{Arutyunov:2010gu}.}
we must take into account that one of the two wings is deformed. The deformed wing contributes
\be
T_{a,0}^{*, {\rm def}} = \frac{(-1)^{a+1}}{Q_{4}^{[1-a]}}
\mathop{\sum_{p=-a+1}^{a-1}}_{\Delta p = 2} Q_{4}^{[p]}.
\ee
For the simplest case of the deformed Konishi which has $L=2$ and spin $N=2$, 
the explicit result for the relevant $Y_{a,0}$ function takes the final form
\be
Y^{*}_{a,0} = -\frac{24576\,a^{2}\,(12u^{2}-a^{2})(12u^{2}+3a^{2}-4)}{(a^{2}+4u^{2})^{3}\,y_{a}\,y_{-a}}
\,\sin^{2}(2\pi\beta)\,g^{6}.
\ee
The residue is
\be
\label{eq:tmp1}
{\rm Res}_{u=i\frac{a}{2}} Y_{a,0}^{*} =
-\frac{216\, i\, a \left(54 a^8-108 a^6+27 a^4+5 a^2-1\right)}{\left(3 a^2-3 a+1\right)^3 \left(3 a^2+3 a+1\right)^3 }
\,\sin^{2}(2\pi\beta)\,g^{6}.
\ee
Summing over $a$ \footnote{It can be checked that this agrees with the sum of the integrals. So, as usual, 
the other poles cancel out in the sum.}, we find the simple result
\be
W_{2}\equiv \sum_{a=1}^{\infty}(-2\,i){\rm Res}_{u=i\frac{a}{2}} Y_{a,0}^{*} =
24\,\sin^{2}(2\pi\beta)\,g^{6}.
\ee
This result can be compared with \cite{Arutyunov:2010gu}. The dispersion relation in \cite{Arutyunov:2010gu} is $\sqrt{1+4g^{2}\sin^{2}\frac{p}{2}}$. Thus their 
coupling is doubled compared to our convention
$g_{ \cite{Arutyunov:2010gu}} = 2g$.
Thus, the result of \cite{Arutyunov:2010gu} which is 
\be
W_{2,  \cite{Arutyunov:2010gu}} = \frac{3}{8}\,\sin^{2}(2\pi\beta)\,g_{ \cite{Arutyunov:2010gu}}^{6},
\ee
is perfectly matched by our calculation.

\subsection{Wrapping correction for deformed twist-2 operators at higher values of $N$}

We have checked that for all the considered $N$
\be
W_{N} = -\frac{1}{\pi}\sum_{a=1}^{\infty}\int_{-\infty}^{\infty}du\,Y_{a,0}^{*} = -2\,i\,
\sum_{a=1}^{\infty}{\rm Res}_{u=i\frac{a}{2}} Y_{a,0}^{*}.
\ee
This is slightly non trivial since it implies that all the extra non-fixed poles do not contribute to the sum.
Now, why is the result rational for $N=2$ ? The reason is that the finite sum over $a$ can be expressed
as a rational function
\ba
\sum_{a=1}^{m}{\rm Res}_{u=i\frac{a}{2}} Y_{a,0}^{*} &=&- \sum_{a=1}^{m}
\frac{216\, i\, a\, \left(54 a^8-108 a^6+27 a^4+5 a^2-1\right)}{\left(3 a^2-3 a+1\right)^3 \left(3 a^2+3 a+1\right)^3 }
\,\sin^{2}(2\pi\beta)\,g^{6} = \nonumber \\
&=& \frac{108\, i\, m \left(3 m^5+9 m^4+14 m^3+13 m^2+6 m+1\right)}{\left(3 m^2+3 m+1\right)^3}\,
\sin^{2}(2\pi\beta)\,g^{6}.\nonumber
\ea
Hence, taking $m\to \infty$ we find 
\be
\sum_{a=1}^{\infty}{\rm Res}_{u=i\frac{a}{2}} Y_{a,0}^{*} =12\,i\,\sin^{2}(2\pi\beta)\,g^{6},\nonumber
\ee
recovering
\be
W_{2} = 24\,\sin^{2}(2\pi\beta)\,g^{6}.
\ee
Remarkably, this works for all even $N$ we have considered. In other words, for all $N$
\be
{\rm Res}_{u=i\frac{a}{2}} Y_{a,0}^{*} = -4\,R_{N}(a)\,\sin^{2}(2\pi\beta)\,g^{6},
\ee
where $R_{N}(a)$ is a rational function such that 
\be
\sum_{a=1}^{m} R_{N}(a) = G_{N}(m),
\ee
where $G$ is another rational function. For instance, at $N=4$, we find
\ba
R_{4}(a) &=& \frac{50 i a A_{4}(a)}{9 \left(35 a^4-70 a^3+85 a^2-50 a+12\right)^3 \left(35 a^4+70 a^3+85 a^2+50 a+12\right)^3}, \\
A_{4}(a) &=& 367653125 a^{20}+3256356250 a^{18}+2399499375 a^{16}-2038277500 a^{14}-4155108125 a^{12} \nonumber \\
&& -6733434750 a^{10}-2048040175 a^8-435524000
   a^6+1012323960 a^4 \nonumber \\
   &&-1910880 a^2-20497536,
\ea
and
\ba
G_{4}(N) &=& -\frac{25 i m B_{4}(m)}{432 \left(35 m^4+70 m^3+85 m^2+50 m+12\right)^3}, \\
B_{4}(m) &=&  986125 m^{11}+5916750 m^{10}+19429725 m^9+42911750 m^8+71073975 m^7+\nonumber\\
&& 91909650 m^6+93405335 m^5+73677330 m^4+43556520 m^3+18264040
   m^2+\nonumber \\
   && 4797216 m+569376
   \ea
   Hence
\be
\sum_{a=1}^{\infty} R_{N}(a) = G_{N}(\infty) = -\frac{25\cdot 986125}{432\cdot35^{3}}\,i = 
-\frac{575}{432}\,i.
\ee
Computing in this way the wrapping for many values of $N$ we find (in the next page) the following list of values of $r_{N}$
appearing in 
\be
W_{N} = -4\,r_{N}\,\,\sin^{2}(2\pi\beta)\,g^{6},
\ee
{
\ba
\label{eq:longlist}
r_{2} &=& -6 \\ 
r_{4} &=& -\frac{575}{216} \nonumber \\ 
r_{6} &=& -\frac{6811}{4500} \nonumber \\ 
r_{8} &=& -\frac{145984913}{148176000} \nonumber \\ 
r_{10} &=& -\frac{3485433677}{5000940000} \nonumber \\ 
r_{12} &=& -\frac{4165633935359}{7987501368000} \nonumber \\ 
r_{14} &=& -\frac{124786385896536779}{307099458846180000} \nonumber \\ 
r_{16} &=& -\frac{61095815015630237}{187184432058624000} \nonumber \\ 
r_{18} &=& -\frac{30864486750624446287}{114954639338002464000} \nonumber \\ 
r_{20} &=& -\frac{33134118246922594651007}{147181789294280328107520} \nonumber \\ 
r_{22} &=& -\frac{821247258526823376973}{4283067942426147114240} \nonumber \\ 
r_{24} &=& -\frac{423296254257078560053417399}{2558229757633583931548851200} \nonumber \\ 
r_{26} &=& -\frac{6752429827603404931071980998039}{46767637756738956248627436000000} \nonumber \\ 
r_{28} &=& -\frac{57655404406777274557039980797413}{453286335180700652871312072000000} \nonumber \\ 
r_{30} &=& -\frac{1338280278251706057435350670024664003}{11844857602202258810226889418580000000} \nonumber \\ 
r_{32} &=& -\frac{10045401921009398169113899909867508536854029}{99368235036141629807876144367566965248000000} \nonumber \\ 
r_{34} &=& -\frac{926715021326134005509987668319993386157239}{10179769257318228260276567820024508320000000} \nonumber \\ 
r_{36} &=& -\frac{25390440807620211250955628419328762124627}{307959406103744720478954992874691008000000} \nonumber \\ 
r_{38} &=& -\frac{48196052500675734519558980197735745588588072947}{642161624325187731270977548626364289361888000000} \nonumber \\ 
r_{40} &=& -\frac{1189640529545796720009723173744957230815007487}{17332297552636645918245008060090804031360000000} \nonumber 
\ea
}

\section{Y-system and twist operators in ABJM}

Another interesting example where a class of twist operators can be found, quite similar to that of \sym,   is the 
 duality between
    Type IIA string theory on ${\rm AdS}_4\times {\mathbb{CP}}^3$
    and planar three-dimensional ${\cal N}=6$ super Chern-Simons theory \cite{Aharony:2008ug}. In this section we will compute wrapping corrections to anomalous dimensions of such operators in this theory.
    
The remarkable integrability properties in this duality have been discussed in many papers, see for instance \cite{AbjmInt}. A Y-system in this context has been initially proposed in \cite{GKV}. This Y-system is defined by the diagram shown in Fig.\ref{Fig:figabjmysys}. That proposal
is only valid    in a large subsector of the theory,
    and a modification has been proposed in \cite{Gromov:2009at,Bombardelli:2009xz} which makes it possible to describe the general case. We do not discuss here in detail the 
    results of    \cite{Gromov:2009at,Bombardelli:2009xz} since we will consider states in the $sl(2)$ sector, for which the results of \cite{GKV} do apply without changes.

\FIGURE[ht]
{\label{Fig:figabjmysys}

    \begin{tabular}{cc}
    \includegraphics[scale=0.6]{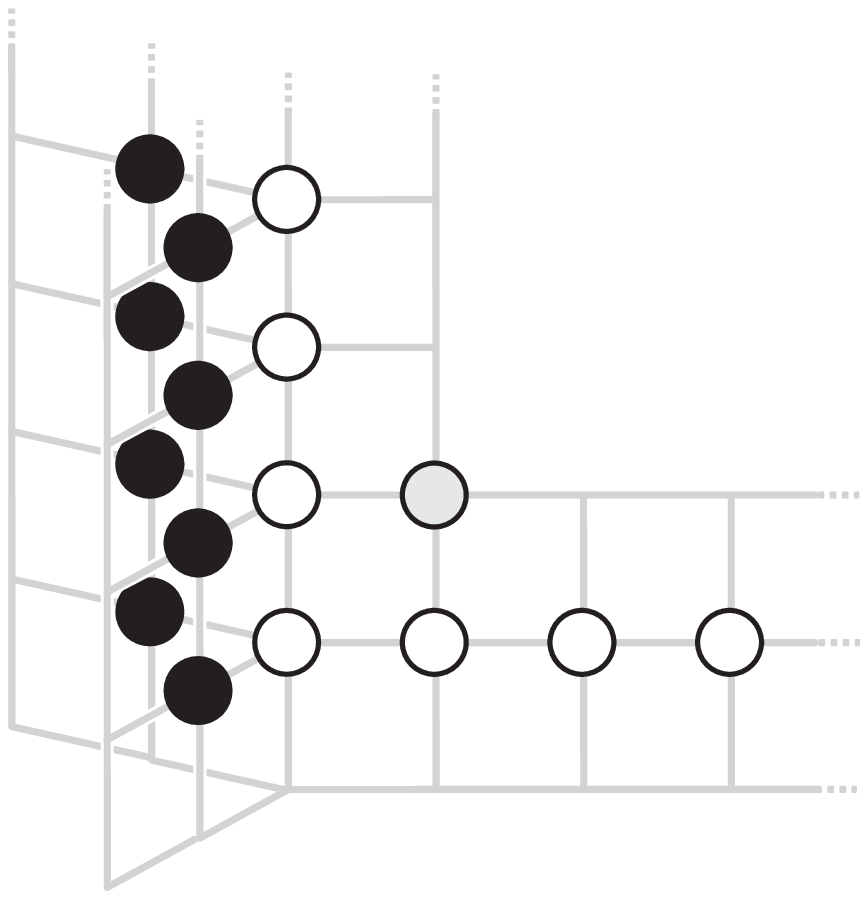}
    \end{tabular}
    \caption{Graphical representation of the Y- and T-systems proposed in \cite{GKV} for ABJM theory.
    Circles in this 3D lattice correspond to Y-functions.}
}

The all-loop Bethe equations for ABJM can be concisely and conveniently summarized by the 
following $\mathfrak{osp}(2,2|6)$ Dynkin diagram (associated with the fermionic $\eta=-1$ grading)
\be
\begin{minipage}{260pt}
\setlength{\unitlength}{1pt}
\small\thicklines
\begin{picture}(160,80)(-60,-50)
%
\put(  0,00){\circle{15}}
\put( -5,-5){\line(1, 1){10}}  
\put( -5, 5){\line(1,-1){10}}  
\put(  0,-15){\makebox(0,0)[t]{$u_1$}} 
\put(  7,00){\line(1,0){26}} 
\put( 40,00){\circle{15}}     
\put( 40,-15){\makebox(0,0)[t]{$u_2$}} 
\put( 47,00){\line(1,0){26}} 
\put( 80,00){\circle{15}}
\put( 75,-5){\line(1, 1){10}}  
\put( 75, 5){\line(1,-1){10}}  
\put( 80,-15){\makebox(0,0)[t]{$u_3$}}  
\put( 88,00){\line(3,2){25}} 
\put(120,20){\circle{15}}
\put(115,15){\line(1, 1){10}}  
\put(115,25){\line(1,-1){10}}  
\put( 88,00){\line(3,-2){25}} 
\put(120,-20){\circle{15}}
\put(115,-25){\line(1, 1){10}}  
\put(115,-15){\line(1,-1){10}}  
\put(150,18){\makebox(0,0)[b]{$u_4$}}
\put(150,-28){\makebox(0,0)[b]{$u_{\overline 4}$}}
\put(118,13){\line(0,-1){25}} 
\put(122,13){\line(0,-1){25}} 
\end{picture}
\end{minipage}
\ee

The energies of magnons in the ABA are again expressed in terms of the Zhukowski variable , but the coupling $g$ in \eq{xdef} is replaced by an 
effective coupling\footnote{Here $\lambda$ denotes the 't Hooft coupling} $h(\lambda)$ (see \cite{MSS,Sh,Bergman:2009zh,McLoughlin:2008he,Bombardelli:2008qd,Leoni:2010tb}), so that
    \be
    \label{xh}
        x + \frac{1}{x} = \frac{u}{h(\lambda)}.
    \ee


We shall consider generalized twist operators in the $\mathfrak{sl}(2)$ sector where we excite symmetrically 
the same number $N$ of $u_4$ and $u_{\overline{4}}$ roots: $u_{4,j}=u_{\bar4,j}$. An explicit description of these states as single trace 
composite operators with length $2L$ can be found in~\cite{Zwiebel:2009vb}. As in the \sym case, 
the integer $L$ can be identified with the twist of the operator.

For such states the Y-functions which correspond to two series of black nodes in Fig.\ref{Fig:figabjmysys} are equal\footnote{in the notation of \cite{GKV}}: $Y^4_{a,0}=Y^{\bar4}_{{a},0}\equiv Y_{a,0}$. In this case the conjectured exact expression for the anomalous dimension \cite{GKV,Gromov:2009at,Bombardelli:2009xz} can be written as
\beq
\label{Eabjm}
    E = 2\sum_{j=1}^{K_4}\epsilon_1^{\rm ph}(u_{4,j})
         + \delta E,\ 
        \delta E=2\sum_{a=1}^{\infty}\int_{-\infty}^{\infty}
        \frac{du}{2\pi i}\frac{\d \epsilon_a^{*}(u)}{\d u}
    \log(1+Y^{*}_{a,0})
\eeq
where $\epsilon_n$ is the energy of a magnon bound state:
\beq
        \epsilon_n (u) \equiv
        \frac{n}{2} + h(\lambda) \left( \frac{i}{x^{[+n]}} - \frac{i}{x^{[-n]}}\right).
\eeq

To compute leading wrapping corrections we use the asymptotic large $L$ solution of the Y-system for ABJM theory, which was obtained in \cite{GKV,Gromov:2009at}. For the $sl(2)$ sector the T-functions in the large $L$ limit can be found from the following generating functional
\ba
\lefteqn{\sum_{a=0}^{\infty} (-1)^{a}\,T_{a,1}\(u+i\tfrac{1-a}{2}\)\,\ddb^{a} = } && \\
&& 
\left(1-\ddb\right)^{-1}\,
\left(1-\(\frac{R_4^{(+)-}}{R_4^{(-)-}}\)^2\,\ddb\right)
\left(1-\(\frac{R_4^{(+)-}}{R_4^{(-)-}}\)^2
\,\ddb\right)\,
\left(1-\(\frac{B_4^{(+)+}}{B_4^{(-)+}}\frac{R_4^{(+)-}}{R_4^{(-)-}}\)^2\ddb\right)^{-1}
\nonumber
\ea
where
\be
R_4^{(\pm)}(u) =  \prod_{\ell=1}^{K_{4}}\left[x(u)-x^{\mp}_{4,\ell}\right], \qquad
B_4^{(\pm)}(u) =  \prod_{\ell=1}^{K_{4}}\left[\frac{1}{x(u)}-x^{\mp}_{4,\ell}\right].
\ee
The $Y_{a,0}$ functions are then given by
\beq
	Y_{a,0}(u)\simeq \(\frac{x^{[-a]}}{x^{[+a]}}\)^L \Phi_a(u) T_{a,1}(u)
\eeq
where the scalar factor is
\beq
	\Phi(u) = \frac{B_4^{(+)+}R_{ 4}^{(-)-}}{B_4^{(-)-}R_{4}^{(+)+}}
	\(\prod_{j=1}^{K_4}\frac{x^+_{4,j}}{x^-_{4,j}}\)\prod_{j=1}^{K_4}\sigma^2(u,u_{4,j}), \ \ 
	\Phi_a(u)=\prod^{\frac{a-1}{2}}_{k=-\frac{a-1}{2}}\Phi(u+ik).
\eeq
Finally, to leading order at weak coupling the wrapping correction obtained from \eq{Eabjm} is
\be
\delta E^{\rm LO}\simeq -\sum_{a=1}^{\infty}\int\frac{du}{\pi} Y_{a,0}^*(u).
\ee
    
\subsection{ABA for twist operators in ABJM}

The asymptotic Bethe equations for states in the $sl(2)$ sector read 
\be
\left(\frac{x^+_k}{x^-_k}\right)^L = -\prod_{j\neq k}^N\frac{u_k-u_j+i}{u_k-u_j-i}\,\left(
\frac{x_k^--x_j^+}{x_k^+-x_j^-}\right)^2\,\sigma^2.
\ee
The only difference compared with ${\cal N}=4$ SYM is the extra minus sign.
 As in the \sym theory, anomalous dimensions will be expanded in powers of $h$
\be
\gamma_{L,N} = L+N+\sum_{\ell\ge 1} \gamma_{L, N}^{(\ell)}\,h^{2\,\ell}.
\ee

\subsection{Four loop wrapping correction to twist-1 operators}

In the twist-1 case, we have the lowest order Bethe Ansatz equations \cite{Beccaria:2009ny}
\be
\frac{u_k+\frac{i}{2}}{u_k-\frac{i}{2}} = -\prod_{j\neq k}^N\frac{u_k-u_j-i}{u_k-u_j+i},\qquad k,j = 1, \dots, N,
\ee
and the Baxter polynomial associated to the ground state  is
\be
Q(u) = {\cal N}\,\prod_{k=1}^N (u-u_k),
\ee
and obeys the equivalent leading order Baxter equation
\be
\left(u+\frac{i}{2}\right)\,Q(u+i)-\left(u-\frac{i}{2}\right)\,Q(u-i) = i\,(2\,N+1)\,Q(u).
\ee
The solution to this recurrence, obeying the polynomiality condition, is 
\be
Q(u) = {}_2F_1\left(\left.\begin{array}{c}
-N, \quad   i\,u+\frac{1}{2} \\
1
\end{array}
\right| 2 \right).
\ee
It follows that the two-loops anomalous dimension can be computed exactly and reads
\be
\gamma^{(2), \ \rm ABA}_{1, N} = \sum_k\frac{2}{u_k^2+\frac{1}{4}} = 4\,\left[S_1(N)-S_{-1}(N)\right].
\ee
Higher order ABA predictions can be found in \cite{Beccaria:2009ny}.
The leading wrapping correction starts at four loops where we split
\be
\gamma^{(4)}_{1, N} = \gamma_{1, N}^{(4), \ \rm ABA}+\gamma_{1, N}^{(4), \ \rm wrapping}.
\ee
The Y-system formulas of  \cite{GKV} allows for the computation of $\gamma_{1, N}^{(4), \ \rm wrapping}$
The result is that it takes the general form 
\be
\gamma_{1, N}^{(4), \ \rm wrapping} = 4\,\left[S_1(N)-S_{-1}(N)\right]\,(r_{N}-2\,\zeta_{2}),
\ee
where the first 20 values of the sequence of rational numbers $r_{N}$ is 
\ba
r_{1} &=& 4, \nonumber \\ 
r_{2} &=& 8/3, \nonumber \\ 
r_{3} &=& 164/45, \nonumber \\ 
r_{4} &=& 932/315, \nonumber \\ 
r_{5} &=& 5552/1575, \nonumber \\ 
r_{6} &=& 159316/51975, \nonumber \\ 
r_{7} &=& 16391656/4729725, \nonumber \\ 
r_{8} &=& 14757016/4729725, \nonumber \\ 
r_{9} &=& 63647092/18555075, \nonumber \\ 
r_{10} &=& 14452397536/4583103525, \nonumber \\ 
r_{11} &=& 171740075876/50414138775, \nonumber \\ 
r_{12} &=& 3682448041828/1159525191825, \nonumber \\ 
r_{13} &=& 13986845259850446488/69870363870782475, \nonumber \\ 
r_{14} &=& 9622892580596/3014765498745, \nonumber \\ 
r_{15} &=& 7381571005683536/2185704986590125, \nonumber \\ 
r_{16} &=& 217095576067044176/67756854584293875, \nonumber \\ 
r_{17} &=& 3878837909713773532/1151866527932995875, \nonumber \\ 
r_{18} &=& 11104533838935576616/3455599583798987625, \nonumber \\ 
r_{19} &=& 8161513390297956691228/2429286507410688300375, \nonumber \\ 
r_{20} &=& 7824894271717769152132/2429286507410688300375. \nonumber 
\ea
Although a closed 
formula was not found for $r_{N}$, it was proved that at large $N$ \cite{Beccaria:2009ny}
\be
\gamma_{1, N}^{(4), \ \rm wrapping} = -\frac{8\,\log N}{N}+\dots
\ee

\subsubsection{Six-loop wrapping correction to twist-2 operators}

The twist-2 case is also discussed in  \cite{Beccaria:2009ny}. We have at lowest order
\be
\left(\frac{u_k+\frac{i}{2}}{u_k-\frac{i}{2}}\right)^{2} = -\prod_{j\neq k}^N\frac{u_k-u_j-i}{u_k-u_j+i},\qquad k,j = 1, \dots, N.
\ee
Now, the Baxter polynomial associated to the ground state, for even $N$,   
obeys the equivalent leading order Baxter equation
\be
\left(u+\frac{i}{2}\right)^{2}\,Q(u+i)-\left(u-\frac{i}{2}\right)^{2}\,Q(u-i) = i\,(2\,N+2)\,u\,Q(u).
\ee
The solution to this recurrence, obeying the polynomiality condition, is 
\be
Q(u) = {}_3F_2\left(\left.\begin{array}{c}
-\frac{N}{2}, \quad   i\,u+\frac{1}{2}, \quad i\,u-\frac{1}{2} \\
1, \quad 1
\end{array}
\right| 1 \right).
\ee
Again, the two-loops anomalous dimension can be computed exactly and reads
\be
\gamma^{(2), \ \rm ABA}_{2, N} = \sum_k\frac{2}{u_k^2+\frac{1}{4}} = 4\,\left[S_1(N)+S_{-1}(N)\right].
\ee
Higher order ABA predictions are also computed in  \cite{Beccaria:2009ny} while the 
the leading wrapping correction in appears at six loops
\be
\gamma^{(6)}_{2, N} = \gamma_{2, N}^{(6), \ \rm ABA} + \gamma^{(6), \rm wrapping}_{2, N}
\ee
has never been discussed. The Y-system provides the following result for the first 15 values
\ba
\gamma^{(6), \rm wrapping}_{2, 2} &=& -\frac{16}{3}+\frac{32 \pi ^2}{9}-\frac{7 \pi ^4}{15}, \\  
\gamma^{(6), \rm wrapping}_{2, 4} &=& -\frac{27}{2}+\frac{48 \pi ^2}{7}-\frac{7 \pi ^4}{10},\nonumber \\  
\gamma^{(6), \rm wrapping}_{2, 6} &=& -\frac{14839}{729}+\frac{15776 \pi ^2}{1701}-\frac{77 \pi ^4}{90},\nonumber \\  
\gamma^{(6), \rm wrapping}_{2, 8} &=& -\frac{1211155}{46656}+\frac{2710000 \pi ^2}{243243}-\frac{35 \pi ^4}{36},\nonumber \\  
\gamma^{(6), \rm wrapping}_{2, 10} &=& -\frac{4465256207}{145800000}+\frac{17747474296 \pi ^2}{1402990875}-\frac{959 \pi ^4}{900},\nonumber \\  
\gamma^{(6), \rm wrapping}_{2, 12} &=& -\frac{31135741}{900000}+\frac{16625130424 \pi ^2}{1195140375}-\frac{343 \pi ^4}{300},\nonumber \\  
\gamma^{(6), \rm wrapping}_{2, 14} &=& -\frac{5177853815453}{136136700000}+\frac{186221441824 \pi ^2}{12422216625}-\frac{121 \pi ^4}{100},\nonumber \\  
\gamma^{(6), \rm wrapping}_{2, 16} &=& -\frac{119246881467979}{2904249600000}+\frac{533851616953168 \pi ^2}{33502718237625}-\frac{761 \pi ^4}{600},\nonumber \\  
\gamma^{(6), \rm wrapping}_{2, 18} &=& -\frac{9171053901749461889}{209602597881600000}+\frac{31508405406517484 \pi ^2}{1878729353479125}-\frac{7129 \pi ^4}{5400},\nonumber \\  
\gamma^{(6), \rm wrapping}_{2, 20} &=& -\frac{11439531709870676071}{247712161132800000}+\frac{1217993024687399444 \pi ^2}{69512986078727625}-\frac{7381 \pi ^4}{5400},\nonumber \\  
\gamma^{(6), \rm wrapping}_{2, 22} &=& -\frac{1930263046069937395188341}{39894291262598572800000}+\frac{2969060731439704190939432 \pi ^2}{163115906021996544626625}-\frac{83711 \pi ^4}{59400},\nonumber \\  
\gamma^{(6), \rm wrapping}_{2, 24} &=& -\frac{77335295702681915396251}{1534395817792252800000}+\frac{8488964619211691016247352 \pi ^2}{450967504884343388085375}-\frac{86021 \pi ^4}{59400},\nonumber \\  
\gamma^{(6), \rm wrapping}_{2, 26} &=& -\frac{506129549311871286088146073871}{9685077248384161620796800000}+\frac{15410719629581936227349212 \pi ^2}{794527352230074116779125}-\frac{1145993 \pi ^4}{772200},\nonumber \\  
\gamma^{(6), \rm wrapping}_{2, 28} &=& -\frac{522794101328166352996044878801}{9685077248384161620796800000}+\frac{385489568436592605572224701676 \pi ^2}{19346197402824463113912844875}+\nonumber \\
&& -\frac{1171733 \pi ^4}{772200},\nonumber \\  
\gamma^{(6), \rm wrapping}_{2, 30} &=& -\frac{10227825075267207999247315567379}{184016467719299070795139200000}+\frac{803678873414383590893537762464 \pi ^2}{39359505060918735300719236125}+\nonumber\\
&&-\frac{1195757 \pi ^4}{772200}.
\nonumber
\ea
A closed formula is definitely unavailable (apart from the pieces $\sim \zeta_{4}$) not to say its large $N$ expansion.

\section{Large spin expansion}
\label{sec:large-spin}

The wrapping contribution is obtained by summing over $a$ the residues 
\be
{\rm Res}_{u=\frac{i}{2}\,a} Y_{1,0}^{*}.
\ee
Once the sum is identified with an explicit function $W_{N}$ of the spin $N$, it is possible to take 
physically interesting limits like large spin $N\to \infty$ to test reciprocity or the analytical continuation 
$N\to -1$ to test BFKL predictions.

In general, such a summation is not available and we propose a novel method to derive it. The idea is to 
obtain closed expressions for the above residues at fixed $a$, expand them in the desired limit, and take
finally the sum over $a$. Potential problems can arise due to the exchange of the limits. Therefore, we test the
method in the undeformed case where a closed formula exists.

\subsection{Undeformed theory}

Let us define $\rho_{a}(N)$ by the residue
\be
{\rm Res}_{u=\frac{i}{2}\,a} Y_{1,0}^{*} = \rho_{a, N}\,g^{8}.
\ee
Janik's formula predicts
\ba
r_{N} &=& -2\,i\,\sum_{a=1}^{\infty} \rho_{a, N} = C_{7, N} +C_{4, N} \zeta_{3}+ C_{2, N} \zeta_{5}, \\
C_{2,N} &=& -640 S_1^2, \\
C_{4,N} &=& -512 S_1^2 S_{-2}, \\
C_{7,N} &=& 256 S_1^2 \left( -S_5+S_{-5}+2S_{4,1}-2S_{3,-2}+2S_{-2,-3}
-4S_{-2,-2,1} \right).
\ea
To inspect the structure of $\rho_{a, N}$, let us begin with the simple case $a=1$. We can work out $\rho_{1, N}$ leaving $Q_{4}$
unspecified, apart from the general requirement $Q_{4}(u)=Q_{4}(-u)$. 
Fixing its normalization by imposing $Q_{4}(\frac{i}{2})=1$,  the result is
\ba
\rho_{1, N} &=& -\frac{4 {Q_4}'\left(\frac{i}{2}\right)^5}{{Q_4}\left(\frac{3 i}{2}\right)}+
\frac{4 {Q_4}'\left(\frac{3 i}{2}\right)
   {Q_4}'\left(\frac{i}{2}\right)^4}{{Q_4}\left(\frac{3 i}{2}\right)^2}
   -\frac{16 i {Q_4}'\left(\frac{i}{2}\right)^4}{{Q_4}\left(\frac{3
   i}{2}\right)}
   -\frac{4 {Q_4}'\left(\frac{3 i}{2}\right)^2 {Q_4}'\left(\frac{i}{2}\right)^3}{{Q_4}\left(\frac{3 i}{2}\right)^3}+
   \nonumber\\
   && \frac{16 i
   {Q_4}'\left(\frac{3 i}{2}\right) {Q_4}'\left(\frac{i}{2}\right)^3}{{Q_4}\left(\frac{3 i}{2}\right)^2}+\frac{40
   {Q_4}'\left(\frac{i}{2}\right)^3}{{Q_4}\left(\frac{3 i}{2}\right)}+\frac{4 {Q_4}'\left(\frac{3 i}{2}\right)^3
   {Q_4}'\left(\frac{i}{2}\right)^2}{{Q_4}\left(\frac{3 i}{2}\right)^4}-\frac{16 i {Q_4}'\left(\frac{3 i}{2}\right)^2
   {Q_4}'\left(\frac{i}{2}\right)^2}{{Q_4}\left(\frac{3 i}{2}\right)^3}+\nonumber \\
   && -\frac{40 {Q_4}'\left(\frac{3 i}{2}\right)
   {Q_4}'\left(\frac{i}{2}\right)^2}{{Q_4}\left(\frac{3 i}{2}\right)^2}+\frac{80 i {Q_4}'\left(\frac{i}{2}\right)^2}{{Q_4}\left(\frac{3
   i}{2}\right)}-\frac{2 {Q_4}^{(3)}\left(\frac{i}{2}\right) {Q_4}'\left(\frac{i}{2}\right)^2}{3 {Q_4}\left(\frac{3 i}{2}\right)}+\nonumber \\
   && \frac{2
   {Q_4}^{(3)}\left(\frac{3 i}{2}\right) {Q_4}'\left(\frac{i}{2}\right)^2}{3 {Q_4}\left(\frac{3 i}{2}\right)^2}+\frac{4 {Q_4}'\left(\frac{i}{2}\right)^3
   {Q_4}''\left(\frac{i}{2}\right)}{{Q_4}\left(\frac{3 i}{2}\right)}+\frac{2 {Q_4}'\left(\frac{i}{2}\right)^3 {Q_4}''\left(\frac{3
   i}{2}\right)}{{Q_4}\left(\frac{3 i}{2}\right)^2}\nonumber \\
   && -\frac{2 {Q_4}'\left(\frac{3 i}{2}\right) {Q_4}'\left(\frac{i}{2}\right)^2
   {Q_4}''\left(\frac{i}{2}\right)}{{Q_4}\left(\frac{3 i}{2}\right)^2}+\frac{8 i {Q_4}'\left(\frac{i}{2}\right)^2
   {Q_4}''\left(\frac{i}{2}\right)}{{Q_4}\left(\frac{3 i}{2}\right)}-\frac{4 {Q_4}'\left(\frac{3 i}{2}\right) {Q_4}'\left(\frac{i}{2}\right)^2
   {Q_4}''\left(\frac{3 i}{2}\right)}{{Q_4}\left(\frac{3 i}{2}\right)^3}+\nonumber \\
   && \frac{8 i {Q_4}'\left(\frac{i}{2}\right)^2 {Q_4}''\left(\frac{3
   i}{2}\right)}{{Q_4}\left(\frac{3 i}{2}\right)^2}.
\ea
Now, the Baxter equation for $Q_{4}(u)$ reads
\be
\left(u+\frac{i}{2}\right)^{2}\, Q_{4}(u+i)+\left(u-\frac{i}{2}\right)^{2} \,Q_{4}(u-i) = 
\left(2u^{2}-N(N+1)-\frac{1}{2}\right)\,Q(u).
\ee
Taking derivatives at $u=\frac{i}{2}$, we can obtain $\rho_{1,N}$ as a function of the first three
derivatives of $Q_{4}$
at $u=\frac{i}{2}$. These are known and can be expressed in terms of harmonic sums~\cite{Beccaria:2009rw}.
Consistently with our normalization, they read
\ba
Q_{4}'\left(\frac{i}{2}\right) &=& -2\,i\,S_{1}, \\
Q_{4}''\left(\frac{i}{2}\right) &=& -4\,(S_{-2}-S_{2}+2S_{1,1}), \\
Q_{4}'''\left(\frac{i}{2}\right) &=& -24\,i\,(S_{-2,1}-S_{1,-2}+S_{1,2}+S_{2,1}-2S_{1,1,1}).
\ea
Hence, we find \footnote{Note that after using the Baxter equation, the terms involving $Q_{4}'''\left(\frac{i}{2}\right)$ drop.}
\ba
\rho_{1,N} &=& 
-\frac{64\, i\, S_1^2}{\left(N^2+N+1\right)^4}
 \left[N^6 \left(-4 S_{1,1}+2 S_1^2-2 S_{-2}+2 S_2+1\right)+ \right. \\
 && \left. N^5 \left(-12 S_{1,1}+6 S_1^2-6 S_{-2}+6 S_2+3\right)+\right. \nonumber \\
 && \left. N^4 \left(-28 S_{1,1}+14 S_1^2+S_1-14
   S_{-2}+14 S_2+7\right)+\right. \nonumber \\
   &&\left. N^3 \left(-36 S_{1,1}+18 S_1^2+2 S_1-18 S_{-2}+18 S_2+9\right)+\right. \nonumber \\
   && \left. N^2 \left(-36 S_{1,1}+18 S_1^2+3 S_1-18 S_{-2}+18 S_2+10\right)+\right. \nonumber \\
   && \left. 2 N \left(-10
   S_{1,1}+5 S_1^2+S_1-5 S_{-2}+5 S_2+3\right)-8 S_{1,1}+4 S_1^2-4 S_{-2}+S_1+4 S_2+5\right].\nonumber
     \ea
The same construction can be repeated for $a>1$. One finds that the Baxter equation allows to reduce the 
calculation of $\rho_{a}(N)$ to a rational function of $N$ and the above three derivatives of $Q_{4}$.

\subsubsection{Large $N$ expansion}

Expanding the harmonic sums $S_{1}, S_{\pm 2}, S_{1,1}, S_{1,1,1}, S_{-2,1}, S_{1,-2}, S_{1,2}, S_{2,1}$
at large $N$ we find the general expansion 
\be
\rho_{a,N} = \frac{i}{2}\sum_{n=2}^{\infty}\sum_{m\ge 0} c_{a}^{n, m}\,\frac{\log^{m}\,\overline N}{N^{n}},
\ee
where $\overline N = N\,e^{\gamma_{E}}$. Remarkably, this is the same form of the large $N$ expansion of
Janik's formula. Hence, one is tempted to sum over $a$ the separate terms in the above expansion.
This is workable because the coefficients $c_{a}^{n, m}$ turn out to be representable by rational functions of $a$
as soon as $a$ is large enough. In the following equations we give their analytic form for all non vanishing cases
with $n\le 5$.

\medskip
\ba
c^{2,2}_{1}         &=& -\frac{64}{3} \left(6+\pi ^2\right), \\
c^{2,2}_{a\ge 2} &=&  -\frac{64 (2 a-1)}{(a-1)^3 a^3},\nonumber 
\ea
\ba
c^{3,2}_{1}         &=& \frac{64}{3} \left(6+\pi ^2\right),\\
c^{3,2}_{a\ge 2} &=& \frac{64 (2 a-1)}{(a-1)^3 a^3}, \nonumber 
\ea
\ba
c^{3,1}_{1}         &=&-\frac{64}{3} \left(6+\pi ^2\right),\\
c^{3,1}_{a\ge 2} &=&  -\frac{64 (2 a-1)}{(a-1)^3 a^3},\nonumber 
\ea
\ba
c^{4,3}_{1}         &=& -128,\\
c^{4,3}_{a\ge 2} &=&  0,\nonumber 
\ea
\ba
c^{4,2}_{1}         &=&-\frac{64 \pi ^2}{3}, \\
c^{4,2}_{2}         &=& \frac{8}{3} \left(21+4 \pi ^2\right),\nonumber\\
c^{4,2}_{a\ge 3} &=&  -\frac{64 \left(3 a^2-6 a+2\right)}{(a-2)^2 (a-1)^2 a^2},\nonumber 
\ea
\ba
c^{4,1}_{1}         &=& \frac{224}{9} \left(6+\pi ^2\right),\\
c^{4,1}_{a\ge 2} &=&  \frac{224 (2 a-1)}{3 (a-1)^3 a^3},\nonumber 
\ea
\ba
c^{4,0}_{1}         &=& -\frac{16}{3} \left(6+\pi ^2\right),\\
c^{4,0}_{a\ge 2} &=&  -\frac{16 (2 a-1)}{(a-1)^3 a^3},\nonumber 
\ea
\ba
c^{5,3}_{1}         &=& 256, \\
c^{5,3}_{a\ge 2} &=&  0,\nonumber 
\ea
\ba
c^{5,2}_{1}         &=& \frac{64}{3} \left(\pi ^2-15\right),\\
c^{5,2}_{2}         &=& -\frac{8}{3} \left(51+8 \pi ^2\right),\nonumber\\
c^{5,2}_{a\ge 3} &=&  \frac{64 \left(6 a^4-20 a^3+25 a^2-16 a+4\right)}{(a-2)^2 (a-1)^3 a^3},\nonumber 
\ea
\ba
c^{5,1}_{1}         &=& -\frac{32}{9} \left(6+7 \pi ^2\right),\\
c^{5,1}_{2}         &=& \frac{4}{3} \left(39+8 \pi ^2\right),\nonumber\\
c^{5,1}_{a\ge 3} &=&  -\frac{32 \left(18 a^4-52 a^3+39 a^2-4\right)}{3 (a-2)^2 (a-1)^3 a^3},\nonumber 
\ea
\ba
c^{5,0}_{1}         &=& \frac{64}{9} \left(6+\pi ^2\right),\\
c^{5,0}_{a\ge 2} &=& \frac{64 (2 a-1)}{3 (a-1)^3 a^3}. \nonumber 
\ea
Summing over $a$, we find the large $N$ expansion
\ba
\label{eq:undef-expansion}
r_{N} &=& \left(-128 \zeta_3-\frac{64 \pi ^2}{3}\right) \frac{\log^{2}\overline N}{N^{2}} +\left(128 \zeta_3+\frac{64 \pi ^2}{3}\right) (\log ^2\overline N- \log 
\overline N)
\, \frac{1}{N^{3}}+\nonumber \\
&& \left[
-128\log^{3}\overline N+\left(32-\frac{32 \pi ^2}{3}\right) \log^2\overline N+\left(\frac{448 \zeta_3}{3}+\frac{224 \pi ^2}{9}\right) \log \overline N-32 \zeta_3-\frac{16 \pi ^2}{3}
\right] \frac{1}{N^{4}}+\nonumber \\
&& \left[
256 \log ^3\overline N+
(-128 \zeta_3-256) \log ^2\overline N+\left(-\frac{64 \zeta_3}{3}+32-\frac{128 \pi ^2}{9}\right) \log \overline N+
\right. \nonumber \\
&& \left. +\frac{128
   \zeta_3}{3}+\frac{64 \pi ^2}{9}
\right] \frac{1}{N^{5}}+\cdots\nonumber \\
\ea
The first line is in perfect agreement with Eq.~(C.2) of \cite{Beccaria:2009vt}. The other contributions can also be checked
to agree with the large $N$ expansion of Janik's formula.

\subsubsection{BFKL poles}

One can immediately see that the exchange of the summation over $a$ and the limit $N\to -1$ (by analytical continuation) {\bf is not legitimate}. The leading pole of $\rho_{a, N}$ as $N\to -1$ is $\sim 1/(N+1)^{4}$, while
the leading pole of Janik's formula is $\sim 1/(N+1)^{7}$.

This is similar to the analitical continuation at $N\to -1$ of 
\be
\zeta(N) = \sum_{a=1}^{\infty}\frac{1}{a^{N}}.
\ee
The separate terms are continued to $a$ and the sum diverges. On the other hand, it is well known that 
$\zeta(-1) = -\frac{1}{12}$.

\begin{figure}[h]
\begin{center}
\includegraphics[scale=0.8]{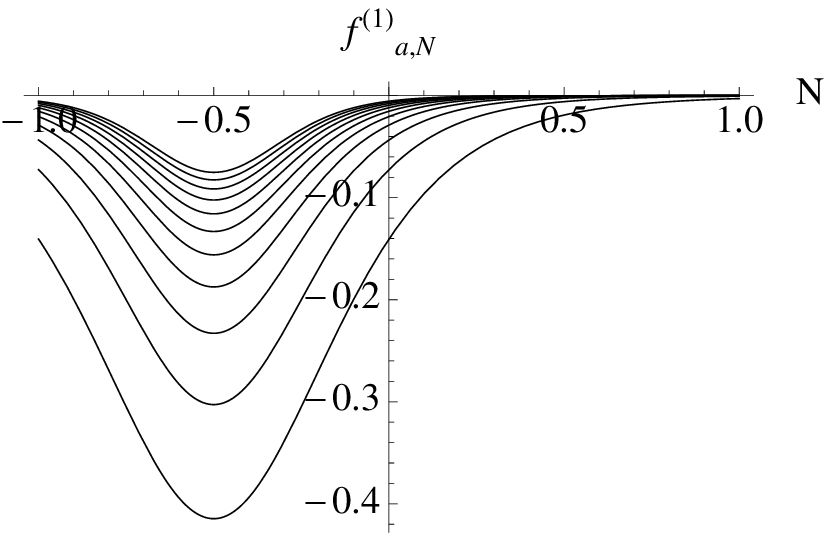}
\includegraphics[scale=0.8]{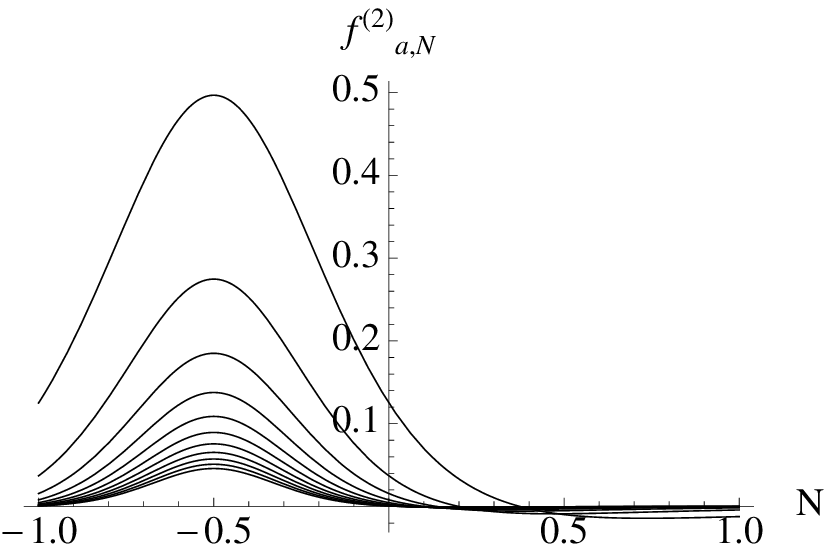} \\
\includegraphics[scale=1]{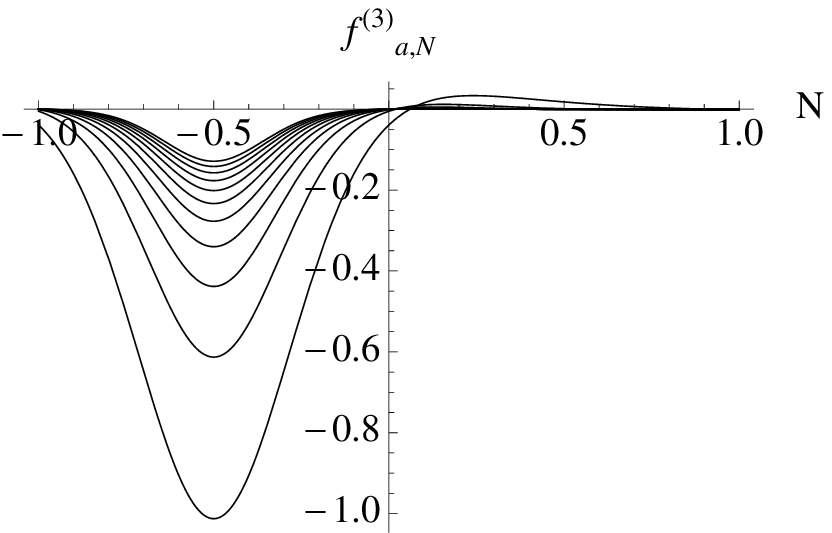}
 \caption{The functions $f^{(1,2,3)}_{a,N}$ as functions of $N$ with $a = 2, \dots, 12$ (moving towards the $N$ axis).}
 \label{fig:f}
 \end{center}
\end{figure}

\begin{figure}[h]
\begin{center}
\includegraphics[scale=1.2]{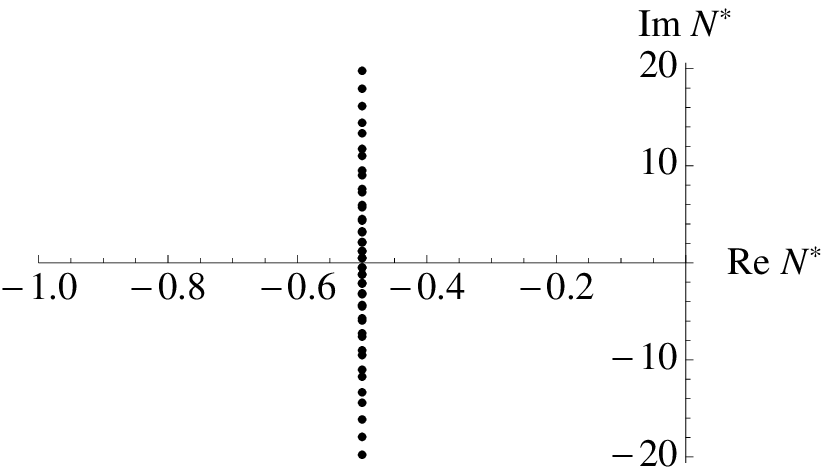}
 \caption{The functions $f^{(1,2,3)}_{a,N}$ as functions of $N$ with $a = 2, \dots, 12$ (moving towards the $N$ axis).}
 \label{fig:sing}
 \end{center}
\end{figure}

In order to understand better what happens, we start from the exact decomposition\footnote{We use $S_{1}^{2} = 
2S_{1,1}-S_{2}$ to eliminate $S_{1,1}$.}
\be
\rho_{a, N} = i\,S_{1}(N)^{3}\,f_{a,N}^{(1)}+ i\,S_{1}(N)^{2}S_{-2}(N)\,f_{a,N}^{(2)}
+ i\,S_{1}(N)^{2}\,f_{a,N}^{(3)},
\ee
where $f^{(1,2,3)}_{a,N}$ are rational functions of $N$ with degrees increasing with $a$. A plot of these coefficient functions is instructive and can be found in Fig.~(\ref{fig:f}). One can check that at fixed positive $N$, the functions 
$f^{(1,2,3)}_{a,N}$ tends rapidly to zero as $a$ increases. Instead, around $N=-1/2$, all of them 
decreases only as $\sim 1/a$ for large $a$. This means that the sum over $a$ diverges across the barrier $N=-1/2$. This is due to accumulation of the (complex) finite $N$ singularities of $\rho_{a,N}$. They have all $Re N^{*}=1/2$ and distribute in the imaginary direction as $a$ increases.
For instance, we show these poles $\{N^{*}\}$ in Fig.~(\ref{fig:sing}) for $a=12$

\subsection{$\beta$-deformed theory}

In the deformed case, we  define $\rho_{a,N}$ by 
\be
{\rm Res}_{u=\frac{i}{2}\,a} Y_{1,0}^{*} = -4\,\rho_{a,N}\,\sin^{2}(2\pi\beta)\,g^{6}.
\ee
Thus, the rational coefficients $r_{N}$ of (\ref{eq:longlist}) are given by the infinite sums
\be
r_{N} = -2\,i\,\sum_{a=1}^{\infty} \rho_{a}.
\ee
Again, we are interested in the large $N$ limit of $r_{N}$. Now, we could not derive a closed formula for $r_{N}$
and therefore we resort to the previous procedure that we tested in the undeformed case.

We begin again with the simple case $a=1$. We can work out $\rho_{1}$ leaving $Q_{4}$
unspecified, apart from the general requirement $Q_{4}(u)=Q_{4}(-u)$. 
Fixing its normalization by imposing $Q_{4}(\frac{i}{2})=1$,  the result is
\ba
\rho_{1, N} &=& 
\frac{Q_4'\left(\frac{i}{2}\right) Q_4''\left(\frac{i}{2}\right)}{Q_4\left(\frac{3
   i}{2}\right)}+
   \frac{Q_4'\left(\frac{i}{2}\right) Q_4''\left(\frac{3 i}{2}\right)}{Q_4\left(\frac{3 i}{2}\right){}^2}-\frac{2
   \left[Q_4'\left(\frac{i}{2}\right)\right]{}^3}{Q_4\left(\frac{3 i}{2}\right)}+
   \frac{2 Q_4'\left(\frac{3 i}{2}\right)
   \left[Q_4'\left(\frac{i}{2}\right)\right]{}^2}{Q_4\left(\frac{3 i}{2}\right){}^2}\nonumber \\
   && -\frac{6 i
   \left[Q_4'\left(\frac{i}{2}\right)\right]{}^2}{Q_4\left(\frac{3 i}{2}\right)}
   -\frac{2 \left[Q_4'\left(\frac{3
   i}{2}\right)\right]{}^2 Q_4'\left(\frac{i}{2}\right)}{Q_4\left(\frac{3 i}{2}\right){}^3}+\frac{6 i Q_4'\left(\frac{3
   i}{2}\right) Q_4'\left(\frac{i}{2}\right)}{Q_4\left(\frac{3 i}{2}\right){}^2}+\frac{12
   Q_4'\left(\frac{i}{2}\right)}{Q_4\left(\frac{3 i}{2}\right)}.\nonumber
\ea
Exploing as before the Baxter equation for $Q_{4}(u)$ we obtain 
\ba
\rho_{1, N} &=& -\frac{4 i S_1}{\left(N^2+N+1\right)^3}
 \left(N^4 \left(-8 S_{1,1}+4 S_1^2-4 S_{-2}+4 S_2+1\right)+\right. \nonumber\\
 &&\left. N^3 \left(-16 S_{1,1}+8 S_1^2-8 S_{-2}+8 S_2+2\right)\right. \nonumber \\
 && \left. -4 N^2 \left(6 S_{1,1}-3
   S_1^2+3 S_{-2}-3 S_2-1\right)+N \left(-16 S_{1,1}+8 S_1^2-8 S_{-2}+8 S_2+3\right)\right.\nonumber\\
   &&\left.  -8 S_{1,1}+4 S_1^2-4 S_{-2}+4 S_2+6\right)
   \ea
The same construction can be repeated for $a>1$. One finds that the Baxter equation allows to reduce the 
calculation of $\rho_{a}(N)$ to a rational function of $N$ and the above three derivatives of $Q_{4}$.

\subsubsection{Large $N$ expansion}

Expanding the harmonic sums $S_{1}, S_{\pm 2}, S_{1,1}, S_{1,1,1}, S_{-2,1}, S_{1,-2}, S_{1,2}, S_{2,1}$
at large $N$ we find the general expansion 
\be
\label{eq:asym}
\rho_{a, N} = \frac{i}{2}\sum_{n=2}^{\infty}\sum_{m=0,1} c_{a}^{n, m}\,\frac{\log^{m}\,\overline N}{N^{n}},
\ee
where $\overline N = N\,e^{\gamma_{E}}$ and  $c_{a}^{2,0}=0$. The other coefficients $c_{a}^{n, m}$ can be represented by rational functions of $a$
as soon as $a$ is large enough. In the following equations we give their analytic form for $n\le 5$.

\medskip
\ba
c^{2,1}_{1}         &=& -\frac{8}{3} \left(3+\pi ^2\right), \\
c^{2,1}_{a\ge 2} &=& \frac{8 (2 a-1)}{(a-1)^2 a^2}, \nonumber 
\ea
\ba
c^{3,1}_{1}         &=& \frac{8}{3} \left(3+\pi ^2\right), \\
c^{3,1}_{a\ge 2} &=& -\frac{8 (2 a-1)}{(a-1)^2 a^2},\nonumber
\ea
\ba
c^{3,0}_{1}         &=& -\frac{4}{3} \left(3+\pi ^2\right), \\
c^{3,0}_{a\ge 2} &=& \frac{4 (2 a-1)}{(a-1)^2 a^2},\nonumber
\ea
\ba
c^{4,1}_{1}         &=& 8, \\
c^{4,1}_{2}         &=& -\frac{2}{3} \left(3+4 \pi ^2\right), \\
c^{4,1}_{a\ge 3} &=& \frac{32 (a-1)}{(a-2)^2 a^2},\nonumber
\ea
\ba
c^{4,0}_{1}         &=& \frac{14}{9} \left(3+\pi ^2\right), \\
c^{4,0}_{a\ge 2} &=& -\frac{14 (2 a-1)}{3 (a-1)^2 a^2},\nonumber
\ea
\ba
c^{5,1}_{1}         &=& -\frac{8}{3} \left(9+\pi ^2\right), \\
c^{5,1}_{2}         &=& \frac{2}{3} \left(15+8 \pi ^2\right), \\
c^{5,1}_{a\ge 3} &=& -\frac{8 \left(6 a^3-15 a^2+12 a-4\right)}{(a-2)^2 (a-1)^2 a^2},\nonumber
\ea
\ba
c^{5,0}_{1}         &=& -\frac{2}{9} \left(\pi ^2-15\right), \\
c^{5,0}_{2}         &=& \frac{1}{6} \left(-3-8 \pi ^2\right), \\
c^{5,0}_{a\ge 3} &=& \frac{2 \left(26 a^3-81 a^2+84 a-28\right)}{3 (a-2)^2 (a-1)^2 a^2}\nonumber, \\
\ea
Summing over $a$, we find the expansion
\ba
\label{eq:expansion}
r_{N} &=& 
-\frac{8\,\pi^{2}}{3}\,\frac{\log\overline N}{N^{2}}
+\left(\frac{8\,\pi^{2}}{3}\,\log\overline N-\frac{4\,\pi^{2}}{3}\right)\,\frac{1}{N^{3}} +\nonumber\\
&& +\left[\left(16-\frac{8\,\pi^{2}}{3}\right)\,\log\overline N+\frac{14\,\pi^{2}}{9} \right]\,\frac{1}{N^{4}} +\\
&& +\left[\left(-32+\frac{8\,\pi^{2}}{3}\right)\,\log\overline N+8-\frac{14\,\pi^{2}}{9} \right]\,\frac{1}{N^{5}}
+\cdots .\nonumber
\ea
Unfortunately, the next term which is $\log\overline N/N^{6}$ has a $c_{a}^{6,1}$ with divergent sum over $a$.
This is a signal of the fact that the expansion (\ref{eq:asym}) is only asymptotic.
This is similar to what happens when one tries to compute, {\em e.g.}, the integral
\be
I(a) = \int_{0}^{\infty}\frac{e^{-at}}{(t+1)^{3}},
\ee
by expanding inside the integral. One obtains
\be
I(a) = \int_{0}^{\infty}dt\left[\frac{1}{(t+1)^3}-\frac{a t}{(t+1)^3}+\frac{a^2 t^2}{2 (t+1)^3}+
\mathcal{O}\left(a^3\right)\right].
\ee
The first two terms give $\frac{1}{2}-\frac{a}{2}$. The third piece diverges. The reason is that $I(a)$ is non analytic
at $a=0$. Indeed, one has
\be
I(a) = \frac{1}{2}-\frac{1}{2}\,a-\frac{1}{2}(\log a-\gamma_{E})\,a^{2}+\mathcal{O}(a^{3}\log a),
\ee
and the convergent terms are correctly reproduced.

To check the accuracy of the asymptotic expansion,
we report in Table (\ref{tab:tab1}) the comparison with the actual values of $r_{N}$ up to $N=40$.

\begin{table}[h]
\begin{center}
\begin{tabular}{lll}
N & $r_{N}$ & expansion (\ref{eq:expansion})  \\
\hline
2 & -6.000000000 & -6.139319435  \\ 
4 & -2.662037037 & -2.665248266  \\ 
6 & -1.513555556 & -1.513877881  \\ 
8 & -0.9852129427 & -0.9852748715  \\ 
10 & -0.6969557077 & -0.6969728432  \\ 
12 & -0.5215190262 & -0.5215250122  \\ 
14 & -0.4063386708 & -0.4063411287  \\ 
16 & -0.3263936768 & -0.3263948131  \\ 
18 & -0.2684927457 & -0.2684933210  \\ 
20 & -0.2251237630 & -0.2251240758  \\ 
22 & -0.1917427577 & -0.1917429379  \\ 
24 & -0.1654645182 & -0.1654646272  \\ 
26 & -0.1443825293 & -0.1443825979  \\ 
28 & -0.1271942257 & -0.1271942703  \\ 
30 & -0.1129840749 & -0.1129841048  \\ 
32 & -0.1010926874 & -0.1010927080  \\ 
34 & -0.09103497318 & -0.09103498770  \\ 
36 & -0.08244736256 & -0.08244737298  \\ 
38 & -0.07505283822 & -0.07505284584  \\ 
40 & -0.06863720900 & -0.06863721466  \\ 
\hline
\end{tabular}
\caption{Actual values of $r_{N}$ compared with the asymptotic expansion  (5.43). }
\label{tab:tab1}
\end{center}
\end{table}

\subsubsection{Gribov-Lipatov Reciprocity}

We can make a further (analytic) non trivial test of the expansion (\ref{eq:expansion}) by checking its generalized
Gribov reciprocity properties (see for instance the recent review \cite{Beccaria:2009vt}). To this aim, we define 
\be
J^{2} = N\,(N+1),
\ee
and expand expansion at large $J$ using the branch
\be
N = \frac{\sqrt{1+4J^{2}}-1}{2}.
\ee
For a generic series in $\log N$ and $1/N$, one should obtain odd powers in $1/J$. Reciprocity is the absence of such
terms. Indeed, we find ($\overline J = J\,e^{\gamma_{E}}$) 

\ba
r_{N} &=& -\frac{4\,\pi^{2}}{3}\,\frac{\log\overline J^{2}}{J^{2}}
+\left(0\cdot\log\overline J^{2}+0\right)\frac{1}{J^{3}}+\\
&&
+\left(-\frac{4\,\pi^{2}}{9}\,\log\overline J^{2}
+8\right)\,\frac{1}{J^{4}}
+\left(0\cdot\log\overline J^{2}+0\right)\frac{1}{J^{5}}+\cdots.\nonumber
\ea
and the terms proportional to odd powers of $1/J$ cancel.

\subsection{ABJM twist-1 operators}

In the ABJM case, we  define $\rho_{a,N}$ by 
\be
{\rm Res}_{u=\frac{i}{2}\,a} Y_{1,0}^{*} = \rho_{a,N}\,h^{4},
\ee
where $h$ is the effective coupling. We repeat the analysis along the lines of the previous cases.

Again, for illustration, we start from the case $a=1$. We can work out $\rho_{1}$ leaving $Q_{4}$
unspecified, apart from the general requirement $Q_{4}(u)=Q_{4}(-u)$. 
Fixing its normalization by imposing $Q_{4}(\frac{i}{2})=1$,  the result (for even N) is
\ba
\rho_{1, N} &=& -\frac{4 i {Q_4}'\left(\frac{i}{2}\right)^2}{{Q_4}\left(\frac{3 i}{2}\right)}+\frac{4 i {Q_4}'\left(\frac{3 i}{2}\right)
   {Q_4}'\left(\frac{i}{2}\right)}{{Q_4}\left(\frac{3 i}{2}\right)^2}+\frac{8 {Q_4}'\left(\frac{i}{2}\right)}{{Q_4}\left(\frac{3 i}{2}\right)}
\ea
The Baxter equation for $Q_{4}(u)$ gives 
\ba
\rho_{1, N} &=& \frac{8 (N+1) }{(2 N+1)^2}\,Q_{4}'\left(\frac{i}{2}\right).
   \ea
This structure~\footnote{A rational function of the spin $N$ times the derivative $Q_{4}'(i/2)$.} is kept also for the other $\rho_{a>1}$. For instance
\be
\rho_{2, N} = \frac{2 \left(8 N^5+20 N^4+26 N^3+15 N^2+2 N-1\right) }{\left(4 N^3+6 N^2+4 N+1\right)^2}
\,Q_{4}'\left(\frac{i}{2}\right)
\ee
We finally remark that one can replace
\be
Q_{4}'\left(\frac{i}{2}\right) = -i\,(S_{1}-S_{-1}).
\ee

\subsubsection{Large $N$ expansion}

Expanding at large $N$ we find the general expansion 
\be
\rho_{a}(N) = \frac{i}{2}\sum_{n=1}^{\infty}\sum_{m=0,1} c_{a}^{n, m}\,\frac{\log^{m}\,(2\overline N)}{N^{n}},
\ee
where $\overline N = N\,e^{\gamma_{E}}$ and
\medskip
\ba
c^{1,1}_{1}         &=&  -4, \\
c^{}_{a\ge 2} &=&  -\frac{4 (-1)^a}{(a-1) a}, \nonumber 
\ea
\ba
c^{2,1}_{1}         &=&  0,\\
c^{2,1}_{a\ge 2} &=&  \frac{2 (-1)^a}{(a-1) a}, \nonumber 
\ea
\ba
c^{3,1}_{1}         &=& 1, \\
c^{3,1}_{2}         &=& -1, \\
c^{3,1}_{a\ge 3} &=& -\frac{(-1)^a (2 a-3)}{(a-2) (a-1)}, \nonumber 
\ea
\ba
c^{3,0}_{1}         &=&  -\frac{2}{3}, \\
c^{3,0}_{a\ge 2} &=&  -\frac{2 (-1)^a}{3 (a-1) a},\nonumber 
\ea
\ba
c^{4,1}_{1}         &=&  -1,\\
c^{4,1}_{2}         &=&  2,\\
c^{4,1}_{a\ge 3} &=&  \frac{(-1)^a (2 a-1) (3 a-4)}{2 (a-2) (a-1) a}, \nonumber 
\ea
\ba
c^{4,0}_{1}         &=&  0,\\
c^{4,0}_{2}         &=&  \frac{1}{6},\\
c^{4,0}_{a\ge 3} &=&  \frac{(-1)^a}{3 (a-1) a},\nonumber 
\ea
Summing over $a$, we find the expansion
\ba
\label{eq:abjm-expansion}
\gamma^{(4), \rm wrapping}_{1, N} &=& -8\log 2\,\frac{\log(2\overline N)}{N}+(4\log 2-2)\,\frac{\log(2\overline N)}{N^{2}}
+\frac{\log(2\overline N)}{N^{3}}+\\
&& -\frac{4}{3}\log 2\,\frac{1}{N^{3}}
+(1-2\log 2)\,\frac{\log(2\overline N)}{N^{4}}+\left(\frac{2}{3}\log 2-\frac{1}{3}\right)\,\frac{1}{N^{4}}+\dots.
\nonumber
\ea
Again, the next term which is $\log(2\overline N)/N^{5}$ has a $c_{a}^{5,1}$ with divergent sum over $a$.
This is a signal of the fact that the expansion (\ref{eq:abjm-expansion}) is only asymptotic.
To check the accuracy of the asymptotic expansion,
we report in Table (\ref{tab:tab2}) the comparison with the actual values of $\frac{1}{h^{4}}\gamma^{(4), \rm wrapping}_{1, N}$ up to $N=20$.
\begin{table}[h]
\begin{center}
\begin{tabular}{lll}
N & $\gamma^{(4), \rm wrapping}_{1, N}$ & expansion (\ref{eq:abjm-expansion})  \\
\hline
2 & -4.985611736 & -4.974203109\\ 
4 & -3.532138400 & -3.531063548\\ 
6 & -2.755402802 & -2.755201899\\ 
8 & -2.277073950 & -2.277016605\\ 
10 & -1.951156397 & -1.951135232\\ 
12 & -1.713589865 & -1.713580601\\ 
14 & -1.532001739 & -1.531997163\\ 
16 & -1.388241842 & -1.388239368\\ 
18 & -1.271317574 & -1.271316139\\ 
20 & -1.174165516 & -1.174164637\\ \hline
\end{tabular}
\caption{Twist-1 in ABJM. Actual values of $\gamma^{(4), \rm wrapping}_{1, N}$ compared with its asymptotic expansion. }
\label{tab:tab2}
\end{center}
\end{table}

\subsection{ABJM twist-2 operators}

Like in the previous case, we  define $\rho_{a,N}$ by 
\be
{\rm Res}_{u=\frac{i}{2}\,a} Y_{1,0}^{*} = \rho_{a,N}\,h^{6},
\ee
where $h$ is the effective coupling. We repeat the analysis along the lines of the previous cases.

Again, for illustration, we start from the case $a=1$. We can work out $\rho_{1}$ leaving $Q_{4}$
unspecified, apart from the general requirement $Q_{4}(u)=Q_{4}(-u)$. 
Fixing its normalization by imposing $Q_{4}(\frac{i}{2})=1$,  the result (for even N) is
\ba
\rho_{1, N} &=& 
-\frac{4 {Q_4}'\left(\frac{i}{2}\right)^3}{{Q_4}\left(\frac{3 i}{2}\right)}+\frac{4
   {Q_4}'\left(\frac{3 i}{2}\right) {Q_4}'\left(\frac{i}{2}\right)^2}{{Q_4}\left(\frac{3
   i}{2}\right)^2}-\frac{12 i {Q_4}'\left(\frac{i}{2}\right)^2}{{Q_4}\left(\frac{3 i}{2}\right)}-\frac{4
   {Q_4}'\left(\frac{3 i}{2}\right)^2 {Q_4}'\left(\frac{i}{2}\right)}{{Q_4}\left(\frac{3
   i}{2}\right)^3}+ \nonumber \\
   && +\frac{12 i {Q_4}'\left(\frac{3 i}{2}\right)
   {Q_4}'\left(\frac{i}{2}\right)}{{Q_4}\left(\frac{3 i}{2}\right)^2}+\frac{24
   {Q_4}'\left(\frac{i}{2}\right)}{{Q_4}\left(\frac{3 i}{2}\right)}+\frac{2
   {Q_4}'\left(\frac{i}{2}\right) {Q_4}''\left(\frac{i}{2}\right)}{{Q_4}\left(\frac{3
   i}{2}\right)}+\\
   && +\frac{2 {Q_4}'\left(\frac{i}{2}\right) {Q_4}''\left(\frac{3
   i}{2}\right)}{{Q_4}\left(\frac{3 i}{2}\right)^2}.\nonumber
\ea
The Baxter equation for $Q_{4}(u)$ gives 
\ba
\rho_{1, N} &=&
-\frac{4 {Q_4}'\left(\frac{i}{2}\right)}{(N+1)^{2}}
 \left[(N+1)\left(
 {Q_4}'\left(\frac{i}{2}\right)^2-{Q_4}''\left(\frac{i}{2}\right)\right)-7N-6
 \right]
 \ea
The first and second derivative of $Q_{4}$ at $u=i/2$ are all we need even for $a>1$.
They can be replaced by the explicit expressions~\footnote{If one prefers, the negative indices can be removed
by writing everything in terms of harmonic sums with argument $N/2$}
\ba
Q_{4}'\left(\frac{i}{2}\right) &=& -i\,(S_{1}+S_{-1}), \\
Q_{4}''\left(\frac{i}{2}\right) &=& -2\,(S_{1,1}+S_{1,-1}+S_{-1,1}+S_{-1,-1}).
\ea

\subsubsection{Large $N$ expansion}

Expanding at large $N$ we find the general expansion 
\be
\rho_{a}(N) = \frac{i}{2}\sum_{n=1}^{\infty}\sum_{m=0,1} c_{a}^{n, m}\,\frac{\log^{m}\,(\frac{1}{2}\overline N)}{N^{n}},
\ee
where $\overline N = N\,e^{\gamma_{E}}$ and

\ba
c^{1,1}_{1}         &=&  \frac{4}{3} \left(\pi ^2-42\right), \\
c^{1,1}_{a\ge 2} &=&  -\frac{8 (-1)^a \left(6 a^2-6 a+1\right)}{(a-1)^2 a^2 (2 a-1)^2}, \nonumber 
\ea
\ba
c^{2,1}_{1}         &=& -\frac{4}{3} (\pi^{2} -36), \\
c^{2,1}_{a\ge 2} &=& \frac{8 (-1)^a \left(6 a^2-6 a+1\right)}{(a-1)^2 a^2 (2 a-1)^2}, \nonumber 
\ea
\ba
c^{2,0}_{1}         &=& \frac{4}{3} \left(\pi ^2-42\right), \\
c^{2,0}_{a\ge 2} &=& -\frac{8 (-1)^a \left(6 a^2-6 a+1\right)}{(a-1)^2 a^2 (2 a-1)^2}, \nonumber 
\ea
\ba
c^{3,1}_{1}         &=& \frac{4}{3} \left(\pi ^2-30\right), \\
c^{3,1}_{2}         &=& -\frac{4}{9} \left(\pi ^2-19\right), \nonumber \\
c^{3,1}_{a\ge 3} &=& -\frac{8 (-1)^a }{(a-2)^2 (a-1)^2 a^2 (2 a-3)^3 (2 a-1)^3}\times \nonumber \\
&&  (48 a^9-336 a^8+780 a^7-68 a^6-3060 a^5+6414 a^4 +\nonumber\\
&& -6242 a^3+3183 a^2-792 a+72), \nonumber 
\ea
\ba
c^{3,0}_{1}         &=& -\frac{8}{9} \left(2 \pi ^2-75\right), \\
c^{3,0}_{a\ge 2} &=& \frac{32 (-1)^a \left(6 a^2-6 a+1\right)}{3 (a-1)^2 a^2 (2 a-1)^2}, \nonumber 
\ea
\ba
c^{4,1}_{1}         &=& -\frac{4}{3} \left(\pi ^2-28\right), \\
c^{4,1}_{2}         &=& \frac{4}{9} \left(3 \pi ^2-70\right),\nonumber \\
c^{4,1}_{a\ge 3} &=& \frac{8 (-1)^a \left(144 a^7-1200 a^6+4260 a^5-8300 a^4+9492 a^3-6282 a^2+2182 a-297\right)}{(a-2)^2 (a-1)^2 (2 a-3)^3 (2 a-1)^3}, \nonumber 
\ea
\ba
c^{4,0}_{1}         &=& \frac{8}{9} \left(2 \pi ^2-63\right), \\
c^{4,0}_{2}         &=& -\frac{2}{27} \left(6 \pi ^2-101\right), \nonumber \\
c^{4,0}_{a\ge 3} &=& -\frac{32 (-1)^a }{3 (a-2)^2 (a-1)^2 a^2 (2 a-3)^3 (2 a-1)^3} \times \nonumber \\
&& (36 a^9-228 a^8+345 a^7+961 a^6-4629 a^5+8001 a^4-7295 a^3+3618 a^2-891 a+81), \nonumber 
\ea
\ba
c^{5,1}_{1}         &=& \frac{4}{3} \left(\pi ^2-34\right), \\
c^{5,1}_{2}         &=& -\frac{4}{27} \left(17 \pi ^2-422\right), \nonumber \\
c^{5,1}_{3}         &=& \frac{2 \left(14727+4000 \pi ^2\right)}{16875}, \nonumber \\
c^{5,1}_{a\ge 4} &=& -\frac{8 (-1)^a }{(a-3)^2 (a-2)^2 (a-1)^2 a^2 (2 a-5)^3 (2 a-3)^4 (2 a-1)^4} \times \nonumber \\
&& (768 a^{17}-10272 a^{16}+11376 a^{15}+704424 a^{14}-7333852 a^{13}+39977984 a^{12} +
\nonumber \\
&& -144003120 a^{11}+370854288 a^{10}-708298920 a^9+1021436220 a^8 +\nonumber \\
&& -1120131812 a^7+933452614 a^6-586698744
   a^5+273890145 a^4+\nonumber \\
   && -92373054 a^3+21347127 a^2-3019680 a+194400), \nonumber 
\ea
\ba
c^{5,0}_{1}         &=& -\frac{8}{5} \left(\pi ^2-27\right), \\
c^{5,0}_{2}         &=& \frac{8}{135} \left(25 \pi ^2-579\right), \nonumber \\
c^{5,0}_{a\ge 3} &=& \frac{16 (-1)^a }{5 (a-2)^2 (a-1)^2 a^2 (2 a-3)^3 (2 a-1)^3}\times \nonumber \\
&& (400 a^9-3312 a^8+11620 a^7-22156 a^6+24292 a^5+\nonumber \\
&& -14614 a^4+3738 a^3+269 a^2-264 a+24). \nonumber 
\ea
Summing over $a$, we find the expansion
\ba
\label{eq:abjm-expansion2}
\gamma^{(6), \rm wrapping}_{2, N} &=& 
\left(
-64 K+\frac{4\pi^{2}}{3}
\right)\,\frac{\log(\frac{1}{2}\overline N)}{N}+
\left(
-8+64 K-\frac{4\pi^{2}}{3}
\right)\,\frac{\log(\frac{1}{2}\overline N)}{N^{2}}+ \nonumber \\
&& + \left(
-64 K+\frac{4\pi^{2}}{3}
\right)\,\frac{1}{N^{2}}+
\left(
\frac{46}{3}-\frac{148}{3} K+\frac{8\pi^{2}}{9}
\right)\,\frac{\log(\frac{1}{2}\overline N)}{N^{3}}+ \nonumber \\
&& +\left(
-8+\frac{256}{3} K-\frac{16\pi^{2}}{9}
\right)\,\frac{1}{N^{3}}+
\left(
-\frac{50}{3}+20 K
\right)\,\frac{\log(\frac{1}{2}\overline N)}{N^{4}}+ \\
&& + 
\left(
18-\frac{212}{3} K+\frac{4\pi^{2}}{3}
\right)\,\frac{1}{N^{4}}+
\left(
\frac{1099}{160}+\frac{2317}{240} K-\frac{32\pi^{2}}{45}
\right)\,\frac{\log(\frac{1}{2}\overline N)}{N^{5}}+ \nonumber \\
&& + \left(
-\frac{196}{9}+\frac{1256}{45} K-\frac{16\pi^{2}}{135}
\right)\,\frac{1}{N^{5}}+\dots , \nonumber
\ea
where $K$ is Catalan's constant defined by 
\be
K = \sum_{n=0}^{\infty}\frac{(-1)^{n}}{(2n+1)^{2}} = 0.915965594177219015054603514932\dots \ .
\ee
To check the accuracy of the asymptotic expansion,
we report in Table (\ref{tab:tab2}) the comparison with the actual values of $\gamma^{(6), \rm wrapping}_{2, N}$ up to $N=30$.
\begin{table}[h]
\begin{center}
\begin{tabular}{lll}
N & $\gamma^{(6), \rm wrapping}_{2, N}$ & expansion (\ref{eq:abjm-expansion2})  \\
\hline
2 & -15.69898239 & -15.82421089\\ 
4 & -14.00907640 & -14.01129166\\ 
6 & -12.15808612 & -12.15830002\\ 
8 & -10.70406871 & -10.70410945\\ 
10 & -9.572736757 & -9.572747981\\ 
12 & -8.674120989 & -8.674124892\\ 
14 & -7.943991150 & -7.943992745\\ 
16 & -7.338680286 & -7.338681019\\ 
18 & -6.828152580 & -6.828152949\\ 
20 & -6.391250615 & -6.391250815\\ 
22 & -6.012693387 & -6.012693501\\ 
24 & -5.681183430 & -5.681183498\\ 
26 & -5.388191030 & -5.388191073\\ 
28 & -5.127154672 & -5.127154700\\ 
30 & -4.892942461 & -4.892942480\\  \hline
\end{tabular}
\caption{Twist-2 in ABJM. Actual values of $\gamma^{(6), \rm wrapping}_{2, N}$ compared with its asymptotic expansion. }
\label{tab:tab3}
\end{center}
\end{table}

\section{Conclusions}
\label{sec:conclusions}

In this paper, we have considered generalized twist operators in $\beta$-deformed \sym and ABJM theory.
We have computed in
several cases the leading wrapping correction at weak coupling from
the Y-systems which had been conjectured for these theories (see
footnote 3).
By exploiting the known one-loop Baxter function of the relevant states, we have obtained 
systematic and accurate large spin expansions for the wrapping effects. In perspective, this has been possible since the 
leading correction is fully determined by the knowledge of the one-loop Baxter polynomial 
for the asymptotic Bethe roots.
When this information is available, the Baxter equation is effectively combined with the Y-system technology 
to control the large spin limit. A remark that we would like to stress is that the wrapping corrections are in a sense 
as simple as the asymptotic leading order energy, at least from the point of view of our investigation. 
Moreover, the method that we have illustrated does not require 
any guesswork to determine the analytical dependence of the wrapping on the spin, something which is not 
available in general. The present work gives a solid foundation to the technical assumption that wrapping corrections
are subleading for the considered operators and theories. An important non-trivial
development will be that of extending our results to the wrapping corrections to excitation over GKP, 
in \sym, or GKP-like strings.

\section*{Acknowledgments}
We thank Nikolay Gromov for very important discussions and clarifications at various stages of this work.
The work of F.L.-M. is supported in part by a grant of the Dynasty Foundation and by the grant RFBR-09-02-00308.




\end{document}